\def\pmb#1{\setbox0=\hbox{#1}%
   \kern-.025em\copy0\kern-\wd0
   \kern.05em\copy0\kern-\wd0
   \kern-0.025em\raise.0433em\box0}
\def\gta{\mathrel{{\lower 3pt\hbox{$\mathchar"218$}}\hskip-8pt
   \raise 2pt\hbox{$\mathchar"13E$}}}
\def\lta{\mathrel{{\lower 3pt\hbox{$\mathchar"218$}}\hskip-8pt
   \raise 2pt\hbox{$\mathchar"13C$}}}
\def\half{{\scriptstyle{1\over2}}}
\def\dagg{\phantom{\dagger}}            

\def\bfnabla{\bf \pmb{\nabla}\/}
\def\bfcalE{\bf \pmb{{\cal E}}\/}

\documentclass[twocolumn,showpacs,preprintnumbers,amsmath,amssymb]{revtex4}

\usepackage{graphicx}
\usepackage{dcolumn}
\usepackage{bm}

\begin{document}

\title{Towards a global theory for the high $\pmb{T_c}\/$ cuprates: \\
Explanation of the puzzling optical properties}

\author{J. Ashkenazi}
 \email{jashkenazi@miami.edu}
\affiliation{%
Physics Department, University of Miami, P.O. Box 248046, Coral
Gables, FL 33124, U.S.A.\\
}%

\date{\today}

\begin{abstract}
A theory has been worked out for the cuprates, which is based on the
major features of their first-principles-derived electronic structure,
including the contribution of a large-$U$ band. Within this theory the
puzzling physics of the cuprates is shown to be a behavior specific of
their structure, within the regime of a Mott transition. The
translational symmetry within the CuO$_2$ planes is disturbed by
dynamical stripe-like inhomogeneities, which provide
quasi-one-dimensional segments where the large-$U$ scenario of
separation between spin and charge is materialized. However, these
charge carriers gain itineracy only due to the coupling with electrons
in the regions where spin and charge are inseparable. Consequently a
two-component scenario is obtained of heavy and light charge carriers,
which are coupled through spin carriers. The theory could explain all
the anomalous properties of the cuprates that were studied by it,
including those observed in transport, tunneling, ARPES, and
neutron-scattering results, the pairing mechanism and its symmetry, the
observed phase diagram, and the occurrence of intrinsic nanoscale
heterogeneity. Here this theory is applied to study a variety of
puzzling optical properties of the cuprates, and again provides a
natural explanation, for each property tested. This includes
``violations'' of the f--sum rule, Tanner's law, Homes' law, Uemura's
law, the behavior of the $n/m^*$ ratio with doping, the behavior in the
heavily underdoped and overdoped regimes, states within the gap and on
its edge, the drop in the $ab$-plane scattering rate below $T^*$ and
$T_c$, the gap-like behavior of the $c$-direction optical conductivity
below $T^*$, and $c$-direction collective modes. 
\end{abstract}

\pacs{71.10.Hf, 71.10.Li, 71.10.Pm, 71.30.+h, 74.20.-z, 74.20.Mn,
74.25.Dw, 74.25.Gz, 74.72.-h} 

\maketitle

\section{Introduction}

The unusual physics of the cuprates, and specifically the occurrence of
high-$T_c$ superconductivity (SC) in them \cite{Muller1}, continues to
be one of the forefront problems in physics. Even though numerous
mechanisms have been proposed to explain this puzzling system ({\it
e.g.}, Refs.~\cite{Brusov, Anderson, Laughlin1, Varma, Lee, Emery,
Kresin, Laughlin, Varma1, Annette, Mihailovic1, Fine}), a consensus has
not been reached yet. The cuprates, as well as related systems of
interest, are characterized by such a complexity that it may make it
unrealistic, at present, to predict the different aspects of their
behavior by a precisely solvable model or numerical calculations. Their
complexity involves single-particle \cite{Pickett,Andersen,Bansil},
many-body \cite{Macridin, Anisimov1, Munoz}, and lattice effects
\cite{Egami,Lanzara1}, as well as the occurrence of nanoscale
inhomogeneity \cite{Bian, Tran1, Muller2, Gorkov, Kapitul, Davis1,
Davis2}. 

Even though approximate numerical calculations, and solutions of simple
models, have been helpful \cite{Macridin, Anisimov1, Munoz, Zaanen1,
Machida, Emery1, Castro, Maki, Dagotto, Scalapino, Auerbach, Norman,
Hanke, Carbotte1, Lichten, Anisimov2, Ovchi, Neto, Eremin2, Eremin1,
Mihailovic2, Markiew}, an approach incorporating results of different
schemes may be necessary for the {\it global} understanding of the
cuprates. In spite that a rigorous analytical or numerical solution of
the incorporated scheme, in its globality, is beyond reach at present,
it provides an insight to the understanding of the physics of the
cuprates, which is missing when models which describe the system
partially are applied. 

Such an approach, including a minimal framework within which different
aspects of the physics of the cuprates could be described correctly, has
been worked out by the author \cite{Ashk01,Ashk03,Ashk04}. Within this 
emerging `global theory' of the cuprates (GTC) their puzzling physics,
including the occurrence of high-$T_c$ SC, can be understood as the
result of a behavior typical of their structure within the regime of a
Mott transition \cite{Lee}. A solution has been studied, corresponding
to a state where translational symmetry within the CuO$_2$ planes is
disturbed by dynamical stripe-like inhomogeneities. Signatures of such
inhomogeneities have been observed, {\it e.g.}, in
Refs.~\cite{Tran1,Kapitul}. They accommodate the competing effects of
hopping and antiferromagnetic (AF) exchange on large-$U$ electrons, and
their existence had been predicted theoretically
\cite{Zaanen1, Machida, Emery1, Castro}. 

These inhomogeneities provide quasi-one-dimensional segments in which
the behavior of the electrons is described in terms of separate carriers
of charge and spin \cite{Anderson}. Since these carriers are strongly
coupled to electrons in the regions where spin and charge are
inseparable, a two-component scenario is obtained of heavy and light
charge carriers, coupled through spin carriers. The speed of the
dynamics of the stripe-like inhomogeneities is self-consistently
determined by the width of spin excitations around the AF wave vector
${\bf Q} = ({\pi \over a} , {\pi \over a})$. Small-width excitations,
and slow dynamics, are obtained in the SC state, and to some degree in
the pseudogap (PG) state \cite{Ashk03,Ashk04}. 

This theory was found to provide a highly-plausible explanation to {\it
all} the anomalous properties of the cuprates that were studied by it.
This includes the systematic anomalous behavior of the resistivity, Hall
constant, and thermoelectric power (TEP) \cite{Ashk01}, and of
spectroscopic anomalies \cite{Ashk03,Ashk04}. Also were explained the
low- and high-energy spin excitations around ${\bf Q}$, including the
neutron-resonance mode \cite{Ashk03,Ashk04}, and its connection to the
peak-dip-hump structure observed in tunneling and ARPES \cite{Ashk04}.
Pairing was shown \cite{Ashk03} to be induced by the energy gain due to
the hopping of pair states perpendicular to the stripe-like
inhomogeneities. Pairing symmetry was predicted to be of the
$d_{x^2-y^2}$ type, but to include features not characteristic of this
symmetry \cite{Ashk03}. 

The phase diagram of the cuprates was found \cite{Ashk04} to result from
an interplay between pairing and coherence within the regime of the Mott
transition. Both pairing and coherence are necessary for SC to occur.
Coherence without pairing results in a metallic Fermi-liquid (FL) state.
Incoherent pairing results in the PG state, consisting of electrons and
localized electron pairs. At low temperatures ($T$), the stripe-like
inhomogeneities partially freeze in the PG state into a glassy
``checkerboard'' structure \cite{Davis1}, and the electrons become
localized. This results in the opening of localization minigaps on the
Fermi surface (FS); they contribute to the PG, and their size determines
the lower doping limit of the SC phase. If SC is suppressed, the
borderline between the FL and the PG states persists down to $T=0$,
where a metal-insulator-transition (MIT) quantum critical point (QCP)
occurs \cite{Boeb,Tallon}. 

A nanoscale heterogeneity was predicted \cite{Ashk04}, especially in the
underdoped (UD) regime, consisting of ``perfectly'' SC regions which
become completely paired for $T \to 0$, and ``PG-like'' SC regions,
where pairing remains partial as in the PG state. A distribution of such
regions of different degrees of pairing has been observed by STM
\cite{Davis2}. This heterogeneity is expected to be intrinsic, and set
in below $T_c$ even in very pure samples. Its scale is larger than that
of the dynamical stripe-like inhomogeneities, which are also intrinsic
and essential for high-$T_c$ SC \cite{Ashk03}. Injection of pairs,
similar to the one occurring in p--n junctions in semiconductors, is
expected in junctions including slices of a cuprate in the SC state and
in the PG state, resulting in the observed ``giant proximity effect''
(GPE) \cite{Bozovic}. 

Similarly to transport, optical properties detect the electrons {\it
within} the crystal, with no transfer of electrons into, or out of it
(as occurs {\it e.g.} in tunneling or ARPES). Thus, even though their
theoretical evaluation involves an integration over the Brillouin zone
(BZ), they still could be very sensitive to fine many-body effects. The
relevance of optical results to the theoretical predictions of Refs.
\cite{Ashk01,Ashk03,Ashk04} has been mentioned there just in passing.
Here it is demonstrated how this GTC naturally explains a variety of
optical results (whichever were tested by it), detailed below, part of
which have been lacking a satisfactory understanding so far. 

The partial Glover--Ferrell--Tinkham sum rule (f--sum rule)
\cite{Kubo1,Tinkham,Norman,Hanke,Marel3}, over the conduction band, is
studied. The GTC is applied to understand observed ``violations'' of the
sum rule through $T_c$, due to the transfer of spectral weight from
energies $\gta 2\;$eV to the vicinity of the Fermi level ($E_{_{\rm
F}}$) \cite{Basov3, Marel1,Bontemps,Homes2}. Also the optical signatures
of $c$-axis collective modes
\cite{Marel3,Marel2,Dulic1,Marel4,Marel5,Dordevic2} are understood. 

The physical interpretation of the optical density to mass ($n/m^*$)
ratio, derived, {\it e.g.}, through the f-sum rule is clarified. The
GTC is shown to explain ``Tanner's law'' \cite{Tanner1},
under which the above ratio is about 4--5 times the contribution to it
from the Drude part of the optical conductivity $\sigma$, which happens
to be just a little greater than the ratio $n_s/m^*$ based on the
low-$T$ superfluid density $n_s$. The approximate factor of four is
connected to the periodicity within the stripe-like inhomogeneities. The
increase in this factor in the heavily UD regime, as well as the
occurrence of a constant effective mass of carriers through the
transition to the AF regime \cite{Basov5}, are also understood. 

The optical quantity $\rho_s = 4 \pi {\rm e}^2 n_s/m^*$ is shown here to
be {\it not} identical with the quantity obtained through the relation
$\rho_s = (c/\lambda)^2$ from measurements of the penetration-depth
$\lambda$ by methods like $\mu SR$ (as has been observed \cite{Tajima}),
due to a difference between the many-body effects on them. In the second
case the GTC predicts \cite{Ashk01,Ashk03,Ashk04} a boomerang-type
behavior in the overdoped (OD) regime, in agreement with experiment
\cite{Niedermayer}. But, as is shown here, no such behavior is expected
for the optical $\rho_s$, also in agreement with experiment
\cite{Timusk1}. 

Within the GTC, the dynamical stripe-like inhomogeneities are
intertwined with low-energy spin excitations around ${\bf Q}$ (including
the resonance mode), which contribute narrow peaks only in the PG and SC
states. It is demonstrated here that their optical signatures have been
observed \cite{Lupi1,Lupi2,Kim,Carbotte2,Timusk} in these states at
energies on the edge of the gap, and within it. 

This theory is also shown to explain the different optical signatures of
the PG in the $c$-direction and in the $ab$-plane. Namely the
observation of a depression, within the PG energy range, of the optical
conductivity $\sigma(\omega)$, in the $c$-direction
\cite{Homes3,Puchkov}, but of the optical scattering rate
$1/\tau(\omega)$ in the $ab$-plane \cite{Puchkov,Basov}. 

The sharp drop in this scattering rate below $T_c$ \cite{Bonn,Puchkov}
is also predicted by the GTC. The existence of the QCP in the
phase diagram \cite{Ashk04} results in marginal-Fermi-liquid (MFL)
\cite{Varma} behavior of the scattering rate above $T_c$, close to the
QCP \cite{Timusk2}, and to critical behavior of optical quantities
\cite{Marel6}. 

``Uemura's law'', under which $T_c \propto \rho_s$ in the UD regime
\cite{Uemura}, has been understood \cite{Ashk01,Ashk03,Ashk04} on the
basis of the pairing phase ``stiffness'' \cite{Emery}. Also the
optically-derived ``Homes' law'' \cite{Homes1}, under which $\rho_s
\simeq 35 \sigma(T_c)T_c$, for cuprates in the entire doping ($x$)
regime, both in the $ab$-plane, and in the $c$-direction, and also for
low-$T_c$ SC's in the dirty limit, is shown to be consistent with the
GTC. The validity of this law in the cuprates is related to the
existence of quantum criticality there \cite{Zaanen}. 

Even though Homes' law is an approximate one (presented in a log--log
scale), its coexistence with Uemura's law in the UD regime implies that
the DC conductivity at $T_c$ varies considerably less with $x$ (in this
regime) than its variation at high temperatures \cite{Takagi}. This
behavior is predicted by the GTC as well, and also are understood
apparent deviations from Uemura's law in the heavily UD regime
\cite{Zuev}. 

\section{Scheme of the Theory \cite{Ashk01,Ashk03,Ashk04}}

\subsection{Large-$U$-limit formalism}

{\it Ab-initio} calculations \cite{Andersen} in the cuprates indicate
that the electrons in the vicinity of $E_{_{\rm F}}$ could be analyzed
in terms of a band (corresponding dominantly to copper and oxygen
orbitals within the CuO$_2$ planes) for which large-$U$-limit
approximations are adequate, which is somewhat hybridized to other bands
for which small-$U$-limit approximations may be suitable. A perturbation
expansion in $U$ is inadequate for the large-$U$ orbitals, and they are
treated by the auxiliary-particles approach \cite{Barnes}. Thus, within
the CuO$_2$ planes, a large-$U$ electron in site $i$ and spin $\sigma$
(which is assigned numbers $\pm 1$, corresponding to $\uparrow$ and
$\downarrow$, respectively) is created by $d_{i\sigma}^{\dagger} =
e_i^{\dagger} s_{i,-\sigma}^{\dagg}$, if it is in the
``upper-Hubbard-band'', and by $d_{i\sigma}^{\prime\dagger} = \sigma
s_{i\sigma}^{\dagger} h_i^{\dagg}$, if it is in a Zhang-Rice-type
``lower-Hubbard-band''. 

Here $e_i^{\dagger}$ and $h_i^{\dagger}$ are creation operators of
``excessions'' and ``holons'', and $s_{i\sigma}^{\dagger}$ are creation
operators of ``spinons''. If either the excessions, or the holons are
ignored, than the large-$U$ band could be treated within the
$t$--$t^{\prime}$--$J$ (or $t$--$t^{\prime}$--$t^{\prime\prime}$--$J$)
model. The auxiliary-particle approach is applied here using the ``slave
fermion'' method \cite{Barnes}, within which the holons/excessions are
fermions and the spinons are bosons. This method had been successful
treating antiferromagnetic (AF) systems, and implies taking good account
of the effect of AF correlations. For rigorous treatment, one should in
principle impose in each site the constraint: $e_i^{\dagger} e_i^{\dagg}
+ h_i^{\dagger} h_i^{\dagg} + \sum_{\sigma} s_{i\sigma}^{\dagger}
s_{i\sigma}^{\dagg} = 1$. 

In order to treat this constraint properly, an auxiliary Hilbert space
is introduced within which a chemical-potential-like Lagrange multiplier
is used to impose the constraint on the average. But since the physical
observables are projected into the physical space as combinations of
Green's functions, whose time evolution is determined by the Hamiltonian
which obeys the constraint rigorously, it is expected that it would not
be violated as long as justifiable approximations were used. 

The dynamics of the stripe-like inhomogeneities is approached
adiabatically, treating them statically with respect to the electrons
dynamics. The striped structure \cite{Tran1} consists of narrow
charged stripes forming antiphase domain walls between wider AF stripes.
Since the spin-charge separation approximation (under which
two-auxiliary-particle spinon--holon/excession Green's functions are
decoupled into single-auxiliary-particle Green's functions) is valid in
one-dimension, it is justified to assume the existence of effective
spinless charge carriers within the narrow charged stripes, but not
within the whole CuO$_2$ plane (as is assumed in RVB theory
\cite{Anderson}). 

\subsection{``Bare'' auxiliary particles}

The spinons are diagonalized by applying the Bogoliubov transformation
for bosons \cite{Ashk94}: $s_{\sigma}^{\dagg}({\bf k}) =
\cosh{(\xi_{\sigma{\bf k}})} \zeta_{\sigma}^{\dagg}({\bf k}) +
\sinh{(\xi_{\sigma{\bf k}})} \zeta_{-\sigma}^{\dagger}(-{\bf k}).$
Spinon states, created by $\zeta_{\sigma}^{\dagger}({\bf k})$, have bare
energies $\epsilon^{\zeta} ({\bf k})$ with a V-shape zero minimum at
${\bf k}={\bf k}_0$. Bose condensation results in an AF order of wave
vector ${\bf Q}=2{\bf k}_0$. Within the lattice BZ there are four
inequivalent possibilities for ${\bf k}_0$: $\pm({\pi \over 2a} , {\pi
\over 2a})$ and $\pm({\pi \over 2a} , -{\pi \over 2a})$, thus
introducing a broken symmetry. One has \cite{Ashk94}: 
\begin{eqnarray}
\cosh{(\xi_{{\bf k}})} &\to& 
\begin{cases} 
+\infty \;,& \text{for ${\bf k} \to {\bf k}_0$,}\\ 1 \;,& \text{for
${\bf k}$ far from ${\bf k}_0$,} 
\end{cases}
\nonumber \\ \sinh{(\xi_{{\bf k}})} &\to& 
\begin{cases}
-\cosh{(\xi_{{\bf k}})} \;,& \text{for ${\bf k} \to {\bf k}_0$,}\\ 0
\;,& \text{for ${\bf k}$ far from ${\bf k}_0$.} 
\end{cases}
\end{eqnarray} 

Holons (excessions) within the charged stripes are referred to as
``stripons''; they carry charge $-{\rm e}$ but no spin, and are created
by fermion operators $p^{\dagger}_{\mu}({\bf k})$. A starting point of
localized stripon states is assumed, due to the fatal effect of
imperfections in the striped structure on itineracy in one dimension.
This assumption is supported by the atomic-scale structure observed
recently by STM \cite{Davis1}, which is consistent with a fluctuating
domino-type two-dimensional arrangement of stripe-like inhomogeneities.
Such a two-dimensional arrangement has been predicted by the author
\cite{Ashk03}. The ${\bf k}$ wave vectors of the stripon states present
${\bf k}$-symmetrized combinations of localized states to be treated in
a perturbation expansion when coupling to the other fields is
considered. 

Away from the charged stripes, creation operators of approximate fermion
basis states of spinon--holon and spinon--excession pairs are
constructed \cite{Ashk01,Ashk03}. Together with the small-$U$ states
they form, within the auxiliary space, a basis to ``quasi-electron''
(QE) states, carrying charge $-{\rm e}$ and spin $\half$, and created by
$q_{\iota\sigma}^{\dagger}({\bf k})$. The bare QE energies
$\epsilon^q_{\iota} ({\bf k})$ form quasi-continuous ranges of bands
within the BZ. 

Atomic doping in the cuprates is taking place in inter-planar layers
(such as in the chains in YBCO) between CuO$_2$ planes, and orbitals of
the doped atoms contribute states close to $E_{_{\rm F}}$. These states
hybridize with the QE bands, and provide charge transfer to the planes
with doping. The QE bands are spanned between the upper and lower
Hubbard bands and the vicinity of $E_{_{\rm F}}$, with more holon-based
states closer to the lower Hubbard band (relevant mainly for ``p-type''
cuprates), and more excession-based states closer to the upper Hubbard
band (relevant mainly for ``n-type'' cuprates). As the doping level $x$
is increasing, more QE states are moving from the Hubbard bands to the
vicinity of $E_{_{\rm F}}$, and the system is moving from the insulating
to the metallic side of the Mott transition regime. 

As was mentioned above, the treatment of the constraint, within the
auxiliary space, introduces an additional chemical-potential-like
Lagrange multiplier. Since the stripons carry only charge, while the
QE's carry both charge and spin, their treatment involves two ``chemical
potentials'', $\mu^q$ and $\mu^p$ (corresponding to QE's and stripons,
respectively) whose values are determined by the correct charge, and
averaged constraint. 

Hopping and hybridization terms introduce strong coupling between the
QE, stripon, and spinon fields, which is expressed by a Hamiltonian term
of the form (for p-type cuprates): 
\begin{eqnarray}
{\cal H}^{\prime} &=& {1 \over \sqrt{N}} \sum_{\iota\lambda\mu\sigma}
\sum_{{\bf k}, {\bf k}^{\prime}} \big\{\sigma
\epsilon^{qp}_{\iota\lambda\mu}(\sigma{\bf k}, \sigma{\bf k}^{\prime})
q_{\iota\sigma}^{\dagger}({\bf k}) p_{\mu}^{\dagg}({\bf k}^{\prime})
\nonumber \\ &\ &\times [\cosh{(\xi_{\lambda,\sigma({\bf k} - {\bf
k}^{\prime})})} \zeta_{\lambda\sigma}^{\dagg}({\bf k} - {\bf
k}^{\prime}) \nonumber \\ &\ &+ \sinh{(\xi_{\lambda,\sigma({\bf k} -
{\bf k}^{\prime})})} \zeta_{\lambda,-\sigma}^{\dagger}({\bf k}^{\prime}
- {\bf k})] + h.c. \big\}, 
\end{eqnarray} 
introducing a vertex between their propagators \cite{Ashk99}). 

The stripe-like inhomogeneities are strongly coupled to the lattice
\cite{Bian}, and it is the presence of stripons which creates the
charged stripes within them. Also the matrix elements in ${\cal
H}^{\prime}$, are sensitive to the atomic positions \cite{Andersen}.
Consequently \cite{Ashk04}, the spinons are renormalized, being
``dressed'' by phonons, and thus carry some lattice distortion (but no
charge) in addition to spin $\half$. Such phonon-dressed spinons are
referred to as ``svivons'', and they replace the spinons in the ${\cal
H}^{\prime}$ vertex. This results in strong coupling between electronic
spin excitations and Cu--O optical phonon modes  \cite{Egami}, and the
existence of an anomalous isotope effect \cite{Lanzara1}. The effect of
spin-lattice coupling has been studied by Eremin {\it et al.} 
\cite{Eremin2}.

\begin{figure}[t] 
\begin{center}
\includegraphics[width=3.25in]{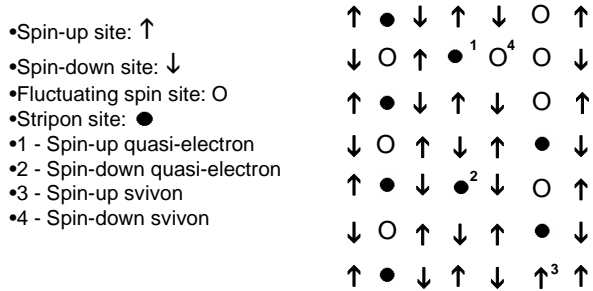}
\end{center}
\caption{An adiabatic ``snapshot'' of a stripe-like inhomogeneity, and
the physical signature of the bare auxiliary particles, within a CuO$_2$
plane.} 
\label{fig1}
\end{figure}

The physical signature of the auxiliary fields, considering only the
large-$U$ band, within the $t$--$t^{\prime}$--$J$ model, is demonstrated
in Fig.~1. An adiabatic ``snapshot'' of a section of a CuO$_2$ plane,
including a stripe-like inhomogeneity, is shown. Within the adiabatic
time scale a site is ``spinless'' either if it is ``charged'', removing
the spinned electron/hole on it (as in ``stripon sites'' in Fig.~1), or
if the spin is fluctuating on a shorter time scale (due to, {\it e.g.},
being in a singlet spin pair). In this description, a site (bare)
stripon excitation represents a transition between these two types of a
spinless site within the charged stripes, a site (bare) svivon
excitation represents a transition between a spinned site and a
fluctuating-spin spinless site, and a site (bare) QE excitation
represents a transition between a spinned site and a charged spinless
site within the AF stripes. 

\subsection{Renormalized auxiliary particles}

The ${\cal H}^{\prime}$ vertex introduces self-energy corrections to the
QE, stripon, and svivon fields \cite{Ashk99}. Since the renormalized
stripon bandwidth is considerably smaller than the QE and svivon
bandwidths, a phase-space argument could be used, as in the Migdal
theorem, to ignore vertex corrections. Self-consistent expressions were
derived \cite{Ashk01,Ashk03} for the self-energy corrections, and
spectral functions $A^q$, $A^p$, and $A^{\zeta}$, for the QE, stripon,
and svivon fields, respectively. The auxiliary-particle energies
$\epsilon$ are renormalized to: $\bar\epsilon = \epsilon +
\Re\Sigma(\bar\epsilon)$, where $\Sigma$ is the self energy. Due to the
quasi-continuous range of QE bands, the bandwidth renormalization is
particularly strong for the stripon energies, resulting in a very small
bandwidth, and limited itineracy due to hopping via intermediary
QE--svivon states. 

The small stripon bandwidth introduces \cite{Ashk01,Ashk03} a low-energy
scale of $\sim 0.02\;$eV, and an apparent ``zero-energy'' non-analytic
behavior of the QE and svivon self energies within a higher energy range
(analyticity is restored in the low-energy range). This behavior results
in QE and svivon scattering rates with a term $\sim \omega$, as in the
MFL approach \cite{Varma}, but also with a constant term, resulting in a
logarithmic singularity in $\Re\Sigma(\omega)$ at $\omega=0$ (which is
truncated by analyticity in the low-energy range). Consequently the QE
self-energy has a kink-like behavior of the renormalized QE energies
$\bar\epsilon^q$ around zero energy \cite{Ashk03}. A typical
renormalization of the svivon energies, around the V-shape zero minimum
of $\epsilon^{\zeta}$ at ${\bf k}_0$, is shown in Fig.~2, where the
major effect is due to the truncated logarithmic singularity in the
svivon self energy. The svivon spectral functions $A^{\zeta}(\omega)$
vary continuously around $\omega=0$, being positive for $\omega>0$ and
negative for $\omega<0$. 

\begin{figure}[t] 
\begin{center}
\includegraphics[width=3.25in]{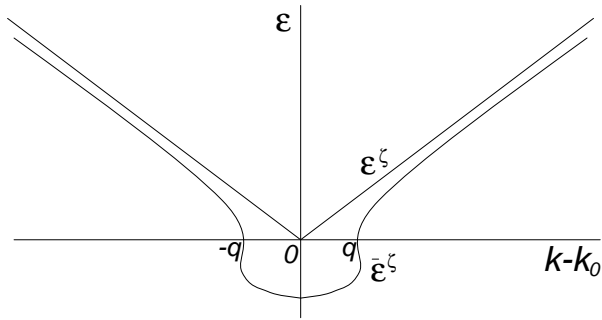}
\end{center}
\caption{A typical self-energy renormalization of the svivon energies
around the minimum at ${\bf k}_0$.} 
\label{fig2}
\end{figure}

The renormalization of the svivon energies changes the physical
signature of their Bose condensation from an AF order to the observed
stripe-like inhomogeneities (including both their spin and lattice
aspects). The structure of $A^{\zeta}$ and $\bar\epsilon^{\zeta}$ around
the minimum at ${\bf k}_0$ (see Fig.~2) determines the structure of the
inhomogeneities. Striped structure of the type shown in Fig.~1 results
from a direction-dependent slope of $\bar\epsilon^{\zeta}({\bf k})$ at
small negative energies. The speed of the dynamics of these
inhomogeneities depends on the linewidth of $\bar\epsilon^{\zeta}$ at
small negative energies, which is large, unless the system is in a
pairing state (thus the SC or PG state), where svivon scattering at such
energies is suppressed due to the gap \cite{Ashk03} (see below).
Consequently, the dynamics of the stripe-like inhomogeneities becomes
sufficiently slow for them to be detected only in a pairing state, in
agreement with experiment. 

For the renormalized svivons, $\cosh{(\xi_{{\bf k}})}$ and
$\sinh{(\xi_{{\bf k}})}$ do not diverge at ${\bf k}_0$, as in Eq.~(1).
But still the values of both of them are large in the
negative-$\bar\epsilon^{\zeta}({\bf k})$ region around this point (see
Fig.~2), and thus this region contributes significantly to processes
involving svivons. By Eq.~(2), the coupling between QE's, stripons, and
svivons of this region, are particularly strong when $\bar\epsilon^q
\simeq \bar\epsilon^p \pm \bar\epsilon^{\zeta}$. On the other hand, a
relatively smaller contribution is obtained to such processes from the
positive-$\bar\epsilon^{\zeta}({\bf k})$ regions, especially when they
are sufficiently away from ${\bf k}_0$, where by Eq.~(1)
$\cosh{(\xi_{{\bf k}})} \simeq 1$, and $\sinh{(\xi_{{\bf k}})} \simeq
0$. 

Since the stripons are based on states in the charged stripe-like
inhomogeneities, which occupy about a quarter of the CuO$_2$ plane
\cite{Tran1} (see Fig.~1), the number of stripon states is about a
quarter of the number of states in the BZ. The ${\bf k}$ values mostly
contributing to these states reflect, on one hand, the structure of the
stripe-like inhomogeneities, and on the other hand, the minimization of
free energy, achieved when they reside mainly at BZ areas where their
coupling to the QE's and svivons, is optimal. This occurs \cite{Ashk03}
for stripon coupling with svivons around ${\bf k}_0$ (see Fig.~2), and
with QE's at BZ areas of highest density of states (DOS) close to
$E_{_{\rm F}}$, which are found in most of the cuprates around the
``antinodal'' points $({\pi \over a} , 0)$ and $(0 , {\pi \over a})$. If
(from its four possibilities) ${\bf k}_0$ were chosen at $({\pi \over
2a} , {\pi \over 2a})$, then the BZ areas in those cuprates, which the
${\bf k}$ values contributing to the stripon states mostly come from,
would be \cite{Ashk03} at about a quarter of the BZ around $\pm{\bf k}^p
= \pm({\pi \over 2a} , -{\pi \over 2a})$. Creation operators
$p^{\dagger}_{\rm e}(\pm{\bf k}^p)$ and $p^{\dagger}_{\rm o}(\pm{\bf
k}^p)$ of stripon states, which are an even and an odd combination of
states at ${\bf k}^p$ and $-{\bf k}^p$, were demonstrated \cite{Ashk04}
to be compatible with the striped structure shown in Fig.~1, in areas
where the stripes are directed along either the $a$ or the $b$
direction. 

\subsection{Hopping-induced pairing}

The electronic structure obtained here for the cuprates provides a
hopping-energy-driven pairing mechanism. Diagrams for such pairing,
based on transitions between pair states of stripons and QE's, through
the exchange of svivons, were presented in Ref.~\cite{Ashk99}. It has
been demonstrated \cite{Ashk03}, within the $t$--$t^{\prime}$--$J$
model, that there is an energy gain in inter-stripe hopping of pairs of
neighboring stripons through intermediary states of pairs of
opposite-spin QE's (where a svivon is exchanged when one pair is
switched to the other), compared to the hopping of two uncorrelated
stripons, through intermediary QE--svivon states (since the intermediary
svivon excitations are avoided). The contribution of orbitals beyond the
$t$--$t^{\prime}$--$J$ model to the QE states results in further gain in
stripon pairing energy, due to both intra-plane and inter-plane pair
hopping. 
 
This pairing scheme provides Eliasherg-type equations, of coupled
stripon and QE pairing order parameters, which can be combined to give
BCS-like equations (though including strong-coupling effects) for both of
them in the second order \cite{Ashk03}. To study these equations, a
domino-type two-dimensional arrangement of stripe-like inhomogeneities,
including crossover between stripe segments directed in the $a$ and the
$b$ directions, was assumed \cite{Ashk03}, and found later to be
consistent with STM results \cite{Davis1}. An overall $d_{x^2-y^2}$-type
pairing symmetry was obtained \cite{Ashk03}, under which pair
correlations are maximal between opposite-spin nearest-neighbor QE
sites, and vanish between same-spin next-nearest-neighbor QE sites (see
Fig.~1). Sign reversal is obtained \cite{Ashk03} for the QE order
parameter through the charged stripes. Thus, the lack of long-range
coherence in the details of the stripe-like inhomogeneities, especially
between different CuO$_2$ planes, results in features different from
those of a simple $d_{x^2-y^2}$-wave pairing (especially when
$c$-direction hopping is involved), as has been observed \cite{Dynes}. 

\subsection{Pairing and coherence}

Within the GTC \cite{Ashk03,Ashk04}, the phase diagram of the cuprates
is largely the consequence of interplay between pairing and coherence
within the regime of a Mott transition. The pairing mechanism, which
depends on the stripe-like inhomogeneities, is stronger when the
AF/stripes effects are stronger, thus closer to the insulating side of
the Mott transition regime. Consequently, the temperature $T_{\rm
pair}$, below which pairing occurs, decreases with the doping level $x$,
as is sketched in the pairing line in Fig.~3. On the other hand, phase
coherence, be it of single electrons or of pairs, requires the
energetic advantage of itineracy around $E_{_{\rm F}}$, which is
easier to achieve closer to the metallic side of the Mott transition
regime. Thus the temperature $T_{\rm coh}$, below which coherence
occurs, increases with $x$, as is sketched in the coherence line in
Fig.~3. Thus $T_c \le \min{(T_{\rm pair}, T_{\rm coh})}$. 

For $x \gta 0.19$, one has $T_{\rm pair} < T_{\rm coh}$, and $T_c$ is
determined by $T_{\rm pair}$. Single-electron coherence, which means the
existence of an FL state, then exists for $T_c < T < T_{\rm coh}$ (see
Fig.~3), as indicated by ARPES results \cite{Ashk03,Ashk04}. Within the
non-FL approach used here, the stripe-like inhomogeneities are treated
adiabatically; but the absence of a pairing gap results in a large
svivon linewidth around ${\bf k}_0$, and thus fast stripes dynamics,
which is consistent with the existence of an FL state below $T_{\rm
coh}$. 

For $x \lta 0.19$, one has $T_{\rm coh} < T_{\rm pair}$, and the
normal-state PG, observed in the cuprates in this regime, is
\cite{Ashk03,Ashk04} (at least partly) a pair-breaking gap at $T_c < T <
T_{\rm pair}$ (see Fig.~3). In this regime $T_{\rm pair}$ is generally
referred to as $T^*$, and $T_c$ is determined by $T_{\rm coh}$. Its
value is of the order of the phase stiffness, estimated through the
treatment of ``classical'' phase fluctuations by the ``X--Y'' model
\cite{Emery}. It is given in this regime by: 
\begin{eqnarray}
k_{_{\rm B}} T_c &\simeq& k_{_{\rm B}} T_{\rm coh} \sim \Big({\hbar
\over {\rm e}}\Big)^2 {a\rho_s \over 16\pi},\\ \rho_s &=& {4 \pi {\rm
e}^2 n_s^* \over m_s^*} = \Big({c \over \lambda}\Big)^2, 
\end{eqnarray}
in agreement with Uemura's law \cite{Uemura} (based on the determination
of the penetration-depth $\lambda$ from $\mu SR$ results). 

\begin{figure}[t] 
\begin{center}
\includegraphics[width=3.25in]{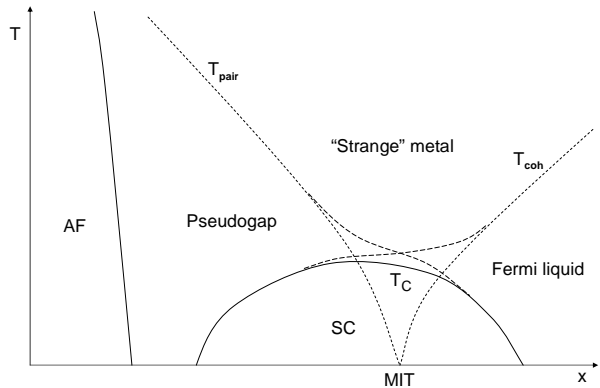}
\end{center}
\caption{A schematic phase diagram for the cuprates. The $T_c$ line is
determined by the pairing line ($T_{\rm pair}$), decreasing with $x$,
and the coherence line ($T_{\rm coh}$), increasing with $x$. Broken
lines should not be regarded as sharp lines (except when $T \to 0$), but
as crossover regimes. The MIT point is a QCP where a metal-insulator
transition occurs at $T=0$ when SC is suppressed.} 
\label{fig3}
\end{figure}

In order to clarify somewhat the phase stiffness energy in Eq.~(3), let
us write it as: $N_s(\hbar K_1)^2/2m_s^*$, where $N_s = n_s^* (Na)^2
a/8\pi^2$ is the average number of pairs in the volume of a slice of
thickness $a/8\pi^2$ around a plane parallel to the CuO$_2$ planes
(assumed to be squares of length $Na$), $m_s^*$ is a pair's mass, and
$K_1 = 2\pi/Na$ is the closest wave-vector point (within a plane) to the
pairs' ${\bf K} = 0$ ground-state point. Thus the phase stiffness is
related to the energy needed (due to lattice discreteness) to excite a
macroscopic number of pairs from the single-pair ground state to the
first excited state. Thus, in the PG state, single-pair states are
occupied (though {\it not} macroscopically) up to energies $\sim
k_{_{\rm B}} T > k_{_{\rm B}} T_{\rm coh}$ (which are a considerable
fraction of the pairs bandwidth). Consequently moderate interactions,
mixing between ${\bf K}$ pair states (such as induced by phonons), are
likely to mix them, resulting in bipolaron-like localized pair states. 

The application of a compressive strain, in this regime, results in the
increase of $a\rho_s$, and thus also of $T_{\rm coh}$ and $T_c$ [see
Eq.~(3)]. And indeed, ARPES results in LSCO thin films \cite{Pavuna}
show that $T_c$ rises under such a strain. As was discussed above, the
increase in $T_{\rm coh}$ is consistent with a move towards the metallic
side of the Mott transition regime, which is reflected in the increase
of the width of the bands around $E_{_{\rm F}}$ under this strain
\cite{Pavuna}. 

When a junction is made, including slices of a cuprate in the SC state
and in the PG state, an injection of pairs is expected to occur between
them, as in p--n junctions in semiconductors. Consequently, the density
of pairs in a range within the PG side of the junction could exceed the
phase-stiffness limit of Eq.~(3) for the occurrence of SC there. This
explains the observation of a GPE in trilayer junctions of cuprate thin
films \cite{Bozovic}, oriented both in the $ab$ plane, and in the $c$
direction. Unlike the regular proximity effect, where the range is
determined by the coherence length, the range of the GPE is determined
by the range where injection of carriers between the SC and PG slices
occurs. This range is not related to the coherence length, and could be
larger by more than an order of magnitude from it, as is observed
\cite{Bozovic}. 

\subsection{Gap equations and heterogeneity}

In order to understand the natures of the SC and PG states, let us
regard the QE and stripon (Bogoliubov) energy bands obtained through
their BCS-like equations \cite{Ashk04}: 
\begin{eqnarray}
E^q_{\pm}({\bf k}) &=& \pm\sqrt{\bar\epsilon^q({\bf k})^2 +
\Delta^q({\bf k})^2}, \\ E^p_{\pm}({\bf k}) &=&
\pm\sqrt{\bar\epsilon^p({\bf k})^2 + \Delta^p({\bf k})^2}.
\end{eqnarray}
$2\Delta^q$ and $2\Delta^p$ are related to the observed pairing gap
\cite{Ashk04}, as will be discussed below. They scale with $T_{\rm
pair}$, approximately according to the BCS factors, with an increase due
to the effect of strong coupling. A $d$-wave pairing factor \cite{Maki}
is relevant for $\Delta^q$, which has its maximum $\Delta^q_{\rm max}$
at the antinodal points. Since the stripons reside in about a quarter of
the BZ around $\pm{\bf k}^p$, $|\Delta^p|$ does not vary much from its
mean value $\bar\Delta^p$, and an $s$-wave pairing factor is relevant
for it. Thus, at low $T$: 
\begin{equation}
2\Delta^q_{\rm max}\gta 4.3 k_{_{\rm B}}T_{\rm pair}, \ \ \ \ \ \
2\bar\Delta^p \gta 3.5 k_{_{\rm B}}T_{\rm pair}. 
\end{equation}

The stripon bandwidth $\omega^p$ has been estimated from TEP results
\cite{Ashk01} (see below) to be $\sim 0.02\;$eV, and by the measured
$T_{\rm pair}$ (see Fig.~3) and the above expression, it is considerably
smaller than $\bar\Delta^p$ in the UD regime, and exceeds its value only
in the heavily OD regime. Consequently, in the UD regime,
$E^p_{\pm}({\bf k}) \simeq \pm|\Delta^p({\bf k})|$, and the Bogoliubov
transformation dictates \cite{Ashk04} an approximate half filling of the
band of the paired stripons. Thus, if all the stripons were paired, the
stripon band would have been approximately half filled (thus $n^p =
\half)$ at low $T$. This is inconsistent with TEP results \cite{Ashk01}
(see below) according to which $n^p > \half$ in the UD regime, and
becomes $\half$ for $x\simeq 0.19$. 

Consequently \cite{Ashk04}, only a part of the stripons could be paired
in the UD regime, while complete QE pairing (except for the nodal and
localized states discussed below) in expected for $T \to 0$. Thus the PG
state consists of {\it both} paired and unpaired stripons. Since an SC
ground state is normally characterized by complete pairing for $T \to
0$, an {\it intrinsically} heterogenous SC state is obtained there
\cite{Ashk04}, with nanoscale perfectly SC regions, where, locally, $n^p
\simeq \half$, and partial-pairing PG-like regions, where $n^p > \half$
locally, but SC still occurs in proximity to the $n^p \simeq \half$
regions. This effect is expected to be weaker, or absent, in the OD
regime, where $\bar\Delta^p$ becomes comparable, and even smaller than
$\omega^p$. 

Such a nanoscale heterogeneity, was indeed observed in STM data in the
SC phase \cite{Davis2}, and it is also supported by optical results
\cite{Dordevic1}. Its features are consistent with the above prediction
\cite{Ashk04} about the existence of perfectly SC regions, and
(especially in the UD regime) partial-pairing PG-like regions. The size
of these regions is comparable to the SC coherence length, so that SC is
maintained also in the partially-paired regions. 

\subsection{Quantum critical point}

The degeneracy of the paired states in a perfectly SC state is
maintained by the dynamics of the stripe-like inhomogeneities
\cite{Ashk04}. Thus the free-energy gain in this state keeps them
dynamical to $T \to 0$. This is not the case in the PG state (and also
within PG-like regions in an heterogenous SC state), where these
inhomogeneities partially freeze at low $T$ into a glassy checkerboard
structure \cite{Davis1,Davis2}. Within this structure one can observe
\cite{Davis1} $a$- and $b$-directed stripe segments, and also $(4a)
\times (4a)$ patterns obtained due to switching between $a$- and
$b$-directed segments, when fluctuations occur between domino-type
two-dimensional arrangements of the inhomogeneities \cite{Ashk03}. 

The formation of this glassy structure is of a similar nature to CDW/SDW
transitions, and orbital effects of the type of the DDW
\cite{Laughlin,Varma1} may also play a role. Energy is gained by
creating partial or complete minigaps for the unpaired carriers (which
may result in an additional superstructure \cite{Davis1}); but even if
there is no real gap within the whole BZ, the unpaired carriers 
would become Anderson-localized due to their low DOS in the low-$T$
disordered glassy structure. And indeed, low-$T$ upturns are observed in
the electrical resistivity in the PG states, either for low $x$, or if
the SC state is suppressed, for $x \lta 0.19$, by applying a magnetic
field \cite{Boeb}, or by doping \cite{Tallon}. When such a suppression
occurs, the pairing and coherence lines (see Fig.~3) meet at $T=0$
around $x\simeq 0.19$, where an MIT occurs \cite{Ashk04} between the FL
metallic phase, and the PG non-metallic phase. 

The $T=0$ MIT point in Fig.~3 satisfies the conditions of a QCP
\cite{Marel6}. The stoichiometry $x$ where it occurs is close to $x_c
\simeq 0.19$, where the fractional stripon occupancy, as was determined
from the TEP results \cite{Ashk01} (see below), is $n^p = \half$. The
existence of the MIT close to this stoichiometry is plausible
\cite{Ashk04}, because for higher doping levels the bare stripons become
too packed within the charged stripes (see Fig.~1), and inter-atomic
Coulomb repulsion between them is likely to destabilize the stripe-like
inhomogeneities in the PG state (though the energy gain in the SC state
helps maintaining them for higher $x$) and stabilize the homogeneous FL
state. 

\section{Previous Applications of the GTC}

\subsection{Electron spectrum}

Spectroscopic measurements (as in tunneling and ARPES) based  on the
transfer of electrons into, or out of, the crystal, are determined by
the electron's spectral function $A_e$, obtained by projecting the
auxiliary spectral functions $A^q$, $A^p$, and $A^{\zeta}$ to the
physical space. Such an expression was derived for $A_e$ \cite{Ashk03},
and it includes a QE ($A^q$) term, and a convoluted stripon--svivon
($A^p A^{\zeta}$) term. From the quasi-continuum of QE bands, only few
bands, which are closely related to those of physical electrons,
contribute ``coherent'' bands, while the other QE bands contribute an
``incoherent'' background to $A_e$. Both the bands and the background
include hybridized $A^q$ and $A^p A^{\zeta}$ contributions, having
widths including \cite{Ashk03} an MFL-type term $\propto\omega$, and a
constant term, in agreement with experiment. $A_e$ is spanned between
the upper and lower Hubbard bands, and the vicinity of $E_{_{\rm F}}$,
and its weight around $E_{_{\rm F}}$ is increasing with $x$. 

Since the stripon states reside (in most cuprates) mainly in a quarter
of the BZ around points $\pm{\bf k}^p$ \cite{Ashk03,Ashk04} (see above),
a significant $A^p A^{\zeta}$ contribution to $A_e$ close to $E_{_{\rm
F}}$ (at energies around $\bar\epsilon^p \pm \bar\epsilon^{\zeta}$) is
obtained, with svivons around their energy minimum at ${\bf k}_0$, in BZ
areas around the antinodal points. Thus the $A^p A^{\zeta}$ contribution
to $A_e$ is {\it not} significant close to ``nodal'' FS crossing points,
in the vicinity of $\pm({\pi \over 2a} , \pm{\pi \over 2a})$, where
$A_e$ is determined primarily by $A^q$. 

Thus the shape of the electron bands around the nodal points is similar
to that of $\bar\epsilon^q$, and the (almost) $T$-independent ``nodal
kink'' observed by ARPES \cite{Lanzara2,Johnson}, closely {\it below}
$E_{_{\rm F}}$, in p-type cuprates, corresponds \cite{Ashk04} to the
$T$-independent kink-like behavior obtained for $\bar\epsilon^q$ there
(see above). The absence of such a kink in ARPES measurements in the
n-type cuprate NCCO \cite{Armitage1} is also consistent with the GTC,
which predicts it to occur there closely {\it above} $E_{_{\rm F}}$
\cite{Ashk04} (and thus out of the range of ARPES). 

On the other hand, the $T$-dependent ``antinodal kink'', observed by
ARPES around the antinodal points \cite{Gromko,Sato,Lanzara1}, where its
major part appears only below $T_c$, is due to the $A^p A^{\zeta}$
contribution to the electron bands there \cite{Ashk04}. Since (see
below) the opening of a pairing gap causes a decrease in the svivon
linewidth around the energy minimum at ${\bf k}_0$ (see Fig.~2), this
contribution narrows down as $T$ is decreased below $T_{\rm pair}$, and
especially below $T_c$ \cite{Ashk04}, as is observed in this kink
\cite{Gromko,Sato,Lanzara1}. 

\subsection{Pair-breaking excitations}

Thus the antinodal kink is a spectroscopic signature of the pairing gap.
Since (in a pairing state) the bandwidth of the Bogoliubov bands $E^p_+$
and $E^p_-$, in Eq.~(6), is small, the convoluted stripon--svivon states
of energies $E^p_+ \pm \bar\epsilon^{\zeta}$ and $E^p_- \pm
\bar\epsilon^{\zeta}$ form \cite{Ashk04} (for svivons around ${\bf
k}_0$) spectral peaks, centered at $E^p_+$ and $E^p_-$, around the
antinodal points. The size of the SC gap is experimentally determined by
the spacing between the closest spectral maxima on its two sides. Thus
it is given by: 
\begin{eqnarray}
2|\Delta^{\rm SC}({\bf k})| &=& 2\min{[|\Delta^q({\bf k})|, E_{\rm
peak}({\bf k})]}, \nonumber \\ E_{\rm peak}({\bf k}) &=& |E^p_{\pm}({\bf
k}\pm{\bf k}_0)|. 
\end{eqnarray}
By Eq.~(7), $\Delta^q_{\rm max} > \bar\Delta^p$; consequently
$\Delta^{\rm SC}$ is determined by $\Delta^q$ around its zeroes at the
nodal points, and by $\Delta^p$ around its maxima at the antinodal
points [where $E_{\rm peak}({\bf k})$ exist]. Actually, since this
convoluted stripon--svivon peak lies on the slope of the QE gap, its
maximum is shifted to an energy slightly above $E_{\rm peak}({\bf k})$.

Since the QE and convoluted stripon--svivon states hybridize with each
other, the states at energies $E^q_{\pm}({\bf k})$, around the antinodal
points, are scattered to stripon--svivon states at energies $E^p_{\pm}
\pm \bar\epsilon^{\zeta}$ of magnitudes above $ E_{\rm peak}({\bf k})$
(see Fig.~2 and the discussion following it). This results in the
widening of the QE coherence peak [due to Eq.~(5)], at the QE gap edge,
to a hump \cite{Ashk04}. 

In the PG state \cite{Ashk04}, the pairs lack phase coherence, and thus
Eq.~(5) does not yield a coherence peak in the QE gap edge. Furthermore,
in this state unpaired convoluted stripon--svivon states exist
\cite{Ashk04} (see discussion above) within the gap, resulting in the
widening of the low-energy svivon states, and thus of $\pm E_{\rm
peak}({\bf k})$, due to scattering. Consequently, the gap-edge
stripon--svivon peak is smeared, and at temperatures well above $T_c$
the PG becomes a depression of width: $2|\Delta^{\rm PG}({\bf k})| =
2|\Delta^q({\bf k})|$ in the DOS \cite{Ashk04}, in agreement with
tunneling results \cite{Renner,Kugler}. 

Thus the pair-breaking excitations in the SC state are characterized by
\cite{Ashk04} a peak-dip-hump structure (on both sides of the gap) in
agreement with tunneling \cite{Renner,Kugler} and ARPES results
\cite{Gromko,Feng1,Borisenko,Janowitz,Plate}. The peak is largely
contributed by the convoluted stripon--svivon states around $E_{\rm
peak}({\bf k})$, the dip results from the sharp descent at the upper
side of this peak, and the hump above them is of the QE gap edge and
other states, widened due to the scattering to stripon--svivon states
above the peak, discussed above. By Eqs.~(7), and (8), $\Delta^{\rm SC}$
and $\Delta^{\rm PG}$ scale with $T_{\rm pair}$, and thus decrease with
$x$, following the pairing line in Fig.~3, as has been observed
\cite{Ashk04}. 

The heterogeneous existence, in an SC state, of perfectly SC and PG-like
partial-pairing regions, discussed above, has been observed by STM
\cite{Davis2} through the distribution in the heights and widths of the
gap-edge peak (widened due to scattering of svivons to unpaired stripons
and QE's). These STM results also show \cite{Davis2} that, unlike the
gap-edge peak, the low energy excitations near the SC gap minimum are
not affected by this heterogeneity. This is consistent with the
prediction \cite{Ashk04} [see Eq.~(8)] that the SC gap is determined,
around its zeroes at the nodal points, by the QE gap $\Delta^q$. The
magnitudes of the QE energies $E^q_{\pm}({\bf k})$ around the nodal
points, are {\it below} the range of $E_{\rm peak}$, and convoluted
stripon--svivon states, which may hybridize with them, correspond to
svivons which are {\it not} at the vicinity of ${\bf k}_0$, and to
energies $E^p_{\pm} \pm \bar\epsilon^{\zeta}$ of magnitudes {\it above}
$E_{\rm peak}$. Thus the hybridization between them is insignificant
(see Fig.~2 and the discussion following it), and the linewidths of the
QE states around the nodal points are small in the SC state, and do {\it
not} vary between the perfectly SC and PG-like regions, as the gap-edge
states do. 

\subsection{Localization gaps, and nodal FS arcs}

As was mentioned above, the formation of the glassy (checkerboard)
structure in the PG state, and the PG-like SC regions, is
self-consistently intertwined with the formation of (at least partial)
minigaps and the localization of the unpaired stripons. Such gaps are
formed there also in QE states around the antinodal points, which are
strongly hybridized with these unpaired stripon states (convoluted with
svivons around  ${\bf k}_0$), and they become localized too. QE states
are extended over a larger range in space than the size of the SC
regions, and those with the above localization gaps can participate in
the pairing process only if these gaps are (approximately) smaller than
their pairing gaps. 

Since the QE bandwidth is much larger than $|\Delta^q|$ [see Eq.~(5)],
they become almost completely paired at low $T$ (also in the PG state),
except for the QE states which have too big localization gaps to
participate in the pairing process, and those at the vicinity of the
nodal points. The minimal doping level $x_0 \simeq 0.05$ \cite{Basov5},
for which SC pairing occurs, is determined by the condition that for $x
\to x_0$ the number of QE states, which have sufficiently small
localization gaps to be coupled to stripon states in the pairing process
\cite{Ashk03}, drops below the minimum necessary for the pairing to
occur. 

Consequently, ARPES measurements in the SC as well as the PG state
\cite{Marshall,Norman1}, show that parts of FS around the antinodal
points disappear, due to the formation of the QE gap (including parts
which are due to localization, and parts which are due to pairing). On
the other hand, arcs of the FS remain around the zero-$\Delta^q$ nodal
points, where unpaired QE states persist at low $T$, and become
localized for $T \to 0$. Such arcs continue to exist also for $x < x_0$
\cite{Yoshida2}. The $x < x_0$ regime is characterized by ``diagonal
stripes'', where the stripon states do not contribute at $E_{_{\rm F}}$
\cite{Yoshida2}, and transport is due to the QE's on the nodal arcs
(with the rest of the FS missing). The modulation caused by the diagonal
stripes, in a direction perpendicular to them, could be the reason for
the minigap observed \cite{Shen} by ARPES in the nodal direction in this
regime. 

Resistivity ($\rho$) measurements through $x=x_0$ \cite{Ando} confirm
the low-$T$ localization on the nodal FS arcs (whether it is Anderson
localization, or due to a minigap), and show a monotonous variation of
$\rho(T)$ with $x$, which is not affected (except for the occurrence of
SC) by the change in the striped structure or the AF transition. This
indicates that the QE's on the nodal FS arcs are hardly affected by the
change in the striped structure. Their hopping is likely to be dominated
by $t^{\prime}$ processes \cite{Ashk03}, which do not disturb the AF
order. One could distinguish between the heavily UD regime \cite{Basov5}
of $x < x_0^{\prime} \simeq 0.09 (> x_0)$, where low $T$ transport is
largely due to QE's on the nodal FS arcs, and the rest of the cuprates
phase diagram (for $x > x_0^{\prime}$), where their role in transport
less important. 

\subsection{Spin excitations and the resonance mode}

Tunneling results \cite{Zasadzinski} show a correlation between the
width of the SC gap-edge peak and the neutron-scattering resonance-mode
energy \cite{Bourges1} $E_{\rm res}$. Such spin excitations are
determined by the imaginary part of the spin susceptibility 
$\chi^{\prime\prime}({\bf k}, \omega)$, and an expression for the
contribution of the large-$U$ orbitals to it has been derived in
Ref.~\cite{Ashk03}. It turns out that large contributions to it are
obtained from double-svivon excitations, when both svivon states are
close to ${\bf k}_0$, and their energies, $\omega_1$ and $\omega_2$,
either have the same sign, and contribute to $\chi^{\prime\prime}({\bf
k}, \omega)$ at $\omega = \pm(\omega_1 + \omega_2)$, or they have
opposite signs, and contribute to it at $\omega = \pm(\omega_1 -
\omega_2)$. This results (see Fig.~2) in two branches of spin
excitations, a low-$\omega$ branch, having a maximum
$-2\bar\epsilon^{\zeta}({\bf k}_0)$ at ${\bf k} = {\bf Q} = 2{\bf k}_0$
(identified as the resonance mode \cite{Bourges1,Reznik}), and a
high-$\omega$ wide branch with an extensive minimum, spreading over the
first branch \cite{Ashk03,Ashk04}. These branches, and also their
linewidths, below and above $T_c$, correspond to neutron-scattering
results \cite{Reznik,Tran2,Hayden,Birgeneau}. [A more quantitative
calculation of $\chi^{\prime\prime}({\bf k}, \omega)$ is in
preparation]. 

The width of the spin excitations is determined \cite{Ashk04} by the
scattering between QE, stripon, and svivon states, which is strong when
$E^q \simeq E^p \pm \bar\epsilon^{\zeta}$, for svivon states close to
${\bf k}_0$ [where the $\cosh{(\xi_{{\bf k}})}$ and $\sinh{(\xi_{{\bf
k}})}$ factors are large  -- see Eqs.~(1) and (2) and the discussion
following Fig.~2]. The existence of a pairing gap limits (especially
below $T_c$) the scattering of the svivon states around ${\bf k}_0$,
resulting in a decrease in their linewidth. Let ${\bf k}_{\rm min}$ be
the points of small svivon linewidth, for which
$\bar\epsilon^{\zeta}({\bf k}_{\rm min})$ is the closest to the energy
minimum $\bar\epsilon^{\zeta}({\bf k}_0)$ (see Fig.~2). Often one has
${\bf k}_{\rm min}={\bf k}_0$, but there are cases, like that of LSCO 
\cite{Ashk04}, where the linewidth of $\bar\epsilon^{\zeta}$ is small
not at ${\bf k}_0$, but at close points ${\bf k}_{\rm min} = {\bf k}_0
\pm {\bf q}$. The resonance mode energy $E_{\rm res}$ is taken here as
$-2 \bar\epsilon^{\zeta}({\bf k}_{\rm min})$, accounting both for the
often observed ``commensurate mode'' at $Q$, and for cases of an
``incommensurate mode'' \cite{Ashk04}, as observed in LSCO at ${\bf Q}
\pm 2{\bf q}$ \cite{Tran3,Wakimoto,Christ}. 

The determination of ${\bf k}_{\rm min}$ is \cite{Ashk04} through the
condition $E_{\rm res} = 2|\bar\epsilon^{\zeta}({\bf k}_{\rm min})| \le
2\tilde\Delta^{\rm SC}$, where $2\tilde\Delta^{\rm SC}$ is somewhat
smaller than the maximal SC gap $2\Delta^{\rm SC}_{\rm max}$. Since
$-\bar\epsilon^{\zeta}({\bf k}_0)$ is zero for an AF (see Fig.~2), its
value (and thus $E_{\rm res}$) is expected to increase with $x$,
distancing from an AF state. However, since its linewidth cannot remain
small if $|\bar\epsilon^{\zeta}({\bf k}_{\rm min})|$ exceeds the value
of $\tilde\Delta^{\rm SC}$, which decreases with $x$, the energy $E_{\rm
res}$ of a {\it sharp} resonance mode is expected \cite{Ashk04} to cross
over from an increase to a decrease with $x$ when it approaches the
value of $2\tilde\Delta^{\rm SC}$, as has been observed \cite{Bourges1}.
This crossover could be followed \cite{Ashk04} by a shift of the
resonance wave vector $2{\bf k}_{\rm min}$ from the AF wave vector ${\bf
Q}$ to incommensurate wave vectors. 

\subsection{The gap-edge peak}

By the above scattering conditions \cite{Ashk04}, svivon energies
$\bar\epsilon^{\zeta}$ have a small linewidth within the range
$|\bar\epsilon^{\zeta}| \le |\bar\epsilon^{\zeta}({\bf k}_{\rm min})|$.
Consequently, the gap-edge peaks of the convoluted stripon--svivon
states of energies $E^p_{\pm} \pm \bar\epsilon^{\zeta}$, centered at
$\pm E_{\rm peak}$ [see Eq.~(8)], have a ``basic'' width: 
\begin{equation}
W_{\rm peak} = -2\bar\epsilon^{\zeta}({\bf k}_{\rm min}) = E_{\rm res}. 
\end{equation}
Additional contributions to the width of this peak come from the svivon
and stripon linewidths, and from the dispersion of $E^p_{\pm}({\bf
k}\pm{\bf k}^{\prime})$ when $\bar\epsilon^{\zeta}({\bf k}_{\rm min})
\lta \bar\epsilon^{\zeta}({\bf k}^{\prime})\lta 0$ (see Fig.~2). This
result explains \cite{Ashk04} the observed correlation (for different
doping levels) between the peak's width, and $E_{\rm res}$
\cite{Zasadzinski}. 

Studies of the pair-breaking excitations in the SC state by ARPES
\cite{Gromko,Feng1,Borisenko,Janowitz,Plate} confirm also the ${\bf k}$
dependence predicted by the GTC \cite{Ashk04}. The convoluted
stripon--svivon states contribute over a range of the BZ around the
antinodal points a single weakly-dispersive gap-edge peak at $E_{\rm
peak}({\bf k})$. Most of the measurements were performed on bilayer
BSCCO, where in addition to this peak there are around the antinodal
points bilayer-split QE bands, the bonding band (BB) and the antibonding
band (AB), contributing two humps around the gap
\cite{Gromko,Feng1,Borisenko}. The AB lies very close to $E_{_{\rm F}}$
on the SC gap edge through the antinodal BZ range, and it almost
overlaps with $E_{\rm peak}({\bf k})$ in the OD regime, where they both
appear as narrow peaks \cite{Gromko,Feng1}. 

The BB, on the other hand, disperses considerably, crossing $E_{_{\rm
F}}$, and contributes a clearly distinguished hump
\cite{Gromko,Feng1,Borisenko}. In the range where the QE BB approaches
$E_{\rm peak}({\bf k})$, the fact that the electron band is formed by
their hybridized contributions results in the appearance of the
antinodal kink \cite{Gromko,Sato,Lanzara1} (discussed above), due to the
narrowing of the peak, as $T$ is lowered below $T_c$. The width of the
hump states depends on the rate of their scattering to stripon--svivon
states. Since the svivons are dressed by phonons, an anomalous isotope
effect is obtained for the width, and thus also for the position of the
hump states \cite{Lanzara1}. 

The peak-dip-hump structure has been observed also in tunneling
\cite{Kugler} and ARPES \cite{Janowitz} measurements in single-layer
BSCO and BSLCO, proving that it is not just the effect of bilayer
splitting \cite{Ashk04}. Recent ARPES measurements in OD single-layer
TBCO \cite{Plate} show the existence of a gap-edge peak, which could be
identified with $E_{\rm peak}({\bf k})$, over a range of the BZ around
the antinodal points. It disperses over a range $\sim 0.02\;$eV, which
is comparable with the SC gap, consistently with the GTC prediction
\cite{Ashk04} for the OD regime. 

Low-temperature ARPES results for the spectral weight within the {\it
sharp} SC gap-edge peak, integrated over the antinodal BZ area
\cite{Feng2,Feng3}, show a maximum for $x\simeq 0.19$. This is expected
by the GTC \cite{Ashk04}, assuming that the integrated spectral weight
is counted within the stripon--svivon peak around $E_{\rm peak}$ in
regions where this peak is sharp. If the number of svivon ${\bf k}$
states contributing to that peak (see Fig.~2) does not vary
significantly with doping in the range of interest, than the measured
integrated peak counts the number of hole-like pair-breaking excitations
of stripons within the $E^p_{-}$ band [see Eq.~(6)], in regions where
the peak is sharp. 

For the intrinsically heterogenous $x\lta 0.19$ regime, discussed above,
this number increases with $x$ because of the increase in the fraction
of space covered by the perfectly SC regions (where the stripon band is
approximately half full and the stripon--svivon states contribute a {\it
sharp} gap-edge peak). For the $x\gta 0.19$ regime there are no PG-like
regions, and the peak is sharp wherever it exists. Thus the measured
integrated peak counts there the number of hole-like pair-breaking
excitations of stripons within the $E^p_{-}$ band, which is decreasing
with the increase of $x$ below half filling of the stripon band
\cite{Ashk04}. Note that the contribution of the QE AB to the ARPES peak
had to be omitted \cite{Feng3} in order to get the decrease of the peak
weight for $x\gta 0.19$, confirming the GTC prediction that this
behavior is due to the stripon--svivon gap-edge peak. 

\subsection{Asymmetry of the tunneling spectrum}

One of the features of the tunneling spectrum in the cuprates
\cite{Davis2,Renner} is its asymmetry with higher DOS for hole- than
particle-excitations. This asymmetry is extending beyond the limit of
the presented spectrum, few tenths of an eV on both sides of $E_{_{\rm
F}}$. Within the GTC, high-$T_c$ in the cuprates is occurring in the
regime of a Mott transition, where the spectral function $A_e(\omega)$
is spanned between the upper and lower Hubbard bands and the vicinity of
$E_{_{\rm F}}$. As was discussed above \cite{Ashk03}, $A_e(\omega)$ has
a large incoherent part, and its magnitude is decreasing when $\omega$
is varied from the Hubbard bands to the energy space between them. 

In p-type cuprates $E_{_{\rm F}}$ is closer to the (Zhang-Rice-type)
lower Hubbard band, and thus $A_e(\omega)$ is descending when $\omega$
is varied from below $E_{_{\rm F}}$ to above it (Anderson and Ong
\cite{Anderson1} demonstrated such a behavior within the $t$--$J$
model), in agreement with the observed asymmetry in the tunneling
spectrum \cite{Davis2,Renner}. In n-type cuprates, $E_{_{\rm F}}$ is
closer to the upper Hubbard band, and thus an {\it opposite} asymmetry
is expected in the tunneling spectrum there (with higher DOS for
particle- than hole-excitations), in agreement with results in n-type
NCCO \cite{Kashiwaya} and ``infinite-layer'' SLCO \cite{Chen}. 

\subsection{Transport properties}

\subsubsection{Expressions within the one-band approximation}

The normal-state transport properties have been a major mystery in the
cuprates. The author has been involved from the start in attempts to
understand the anomalous TEP, resistivity, and Hall constant
\cite{Fisher,Bar-Ad}. Their correct treatment, within the GTC, though
not including the effects of the PG, was first presented in
Ref.~\cite{Ashk01}. Even though, the dynamics of the stripe-like
inhomogeneities is fast above $T_{\rm pair}$, their adiabatic treatment
is still expected to be a reasonable approximation for the evaluation of
the transport properties. Their derivation, within the $ab$ plane, is
based on linear-response theory, A condition used is that the direct
current (DC) ${\bf j}$ could be expressed as a sum of QE (${\bf j}^q$)
and stripon (${\bf j}^p$) terms which are proportional to each other
with an approximately $T$-independent proportionality factor $\alpha$.
Thus: 
\begin{equation}
{\bf j} = {\bf j}^q + {\bf j}^p \cong {{\bf j}^q \over 1-\alpha} \cong
{{\bf j}^p \over \alpha}. 
\end{equation}
As is discussed further below, in relation to optical conductivity, this
condition is a consequence of the assumption that the contribution of
the CuO$_2$ planes to transport is derived, dominantly, from {\it one}
electron band of the homogeneous planes. The formation of stripe-like
inhomogeneities results in separate contributions of QE's and stripons,
but the stripes dynamics results in DC based on the band of the
averaged homogeneous planes. The coefficient $\alpha$ depends on the
inhomogeneous structure, which determines how ${\bf j}$ is split between
${\bf j}^q$ and ${\bf j}^p$. 

Transport can be treated using the separate QE and stripon chemical
potentials, $\mu^q$ and $\mu^p$ (due to the large-$U$ constraint),
discussed above. When an electric field ${\bf E}$ is applied, Eq.~(10)
is satisfied by the formation of gradients ${\bfnabla}\mu^q$ and
${\bfnabla}\mu^p$ \cite{Ashk01}, where the homogeneity of charge
neutrality imposes: 
\begin{equation}
{\partial n^q_e \over \partial \mu^q} {\bfnabla}\mu^q + {\partial n^p_e
\over \partial \mu^p} {\bfnabla}\mu^p =0. 
\end{equation}
Here $n^q_e$ and $n^p_e$ are the contributions of QE and stripon states
to the electrons occupation. Because of the small stripon bandwidth, it
is convenient to introduce \cite{Ashk01}: 
\begin{equation}
N^q_e \equiv {\partial n^q_e \over \partial \mu^q}, \ \ \ \ \ M^p_e(T)
\equiv T {\partial n^p_e \over \partial \mu^p}, 
\end{equation}
where $N^q_e$ is almost temperature independent, and $M^p_e(T)$
saturates at temperatures above the stripon bandwidth to a constant
$n^p(1-n^p)$ \cite{Ashk01}. 

The chemical potential gradients introduce  ``chemical fields'',
resulting in different effective fields for QE's and stripons
\cite{Ashk01}: 
\begin{equation}
{\bfcalE}^q={\bf E} + {\bfnabla}\mu^q / {\rm e}, \ \ \ \ \ 
{\bfcalE}^p={\bf E} + {\bfnabla}\mu^p / {\rm e}, 
\end{equation}
and by Eqs.~(11), (12), and (13), one gets: 
\begin{equation}
{\bf E}={M^p_e(T) {\bfcalE}^p + N^q_e T {\bfcalE}^q \over M^p_e(T) +
N^q_eT}. 
\end{equation}
Effective QE and stripon conductivities $\sigma^q$ and
$\sigma^p$ are introduced through the expressions:
\begin{equation}
{\bfcalE}^q =  {\bf j}/\sigma^q(T), \ \ \ \ \ {\bfcalE}^p = {\bf
j}/\sigma^p(T), 
\end{equation}
and related expressions are used \cite{Ashk01} to introduce effective QE
and stripon Hall numbers $n^q_{_{\rm H}}$ and $n^p_{_{\rm H}}$, and
TEP's $S^q$ and $S^p$. Using Eq.~(14), the following expressions are
then obtained \cite{Ashk01} for the electrical conductivity ($\sigma$),
Hall number ($n_{_{\rm H}}$), and TEP ($S$): 
\begin{eqnarray}
\sigma &=& {M^p_e(T) + N^q_eT \over M^p_e(T) / \sigma^p(T) +  N^q_e T /
\sigma^q(T)}, \\ 
n_{_{\rm H}} &=& {M^p_e(T) + N^q_eT \over M^p_e(T) / n^p_{_{\rm H}} +
N^q_e T / n^q_{_{\rm H}}}, \\ 
S &=& {M^p_e(T) S^p(T) + N^q_e T S^q(T) \over M^p_e(T) + N^q_eT}. 
\end{eqnarray}

At low $T$, for $x \gg x_0^{\prime}$, the value of $M^p_e(T)$ is
generally about an order of magnitude greater than $N^q_eT$.
Consequently, by Eqs.~(11) and (12), ${\bfnabla}\mu^p\cong 0$ and
${\bfnabla}\mu^q \cong -{\rm e}{\bf E}$. Thus, by Eq.~(13), ${\bfcalE}^p
\cong {\bf E}$, ${\bfcalE}^q \cong 0$, and by Eqs.~(16), (17) and (18),
the transport coefficients are dominantly stripon-like at low $T$, both
above and below $T_c$, and a crossover to a QE-like behavior is
occurring when $T$ is increased through the stability temperature
regime. When pairing gaps exist, $\mu^q$ and $\mu^p$ are within these
gaps, and their existence results in some decrease in the low-$T$ value
of $M^p_e(T)$ in Eq.~(12), resulting in some increase in the QE
contribution to transport at low $T$. The assumption that $M^p_e(T) \gg
N^q_eT$, at low $T$, breaks down when $x$ approaches $x_0^{\prime}$
(towards the heavily UD regime), where the PG and SC gap are large. As
was discussed above, low-$T$ transport for $x < x_0^{\prime}$ is largely
due to QE's, which are on the nodal FS arcs, for the unpaired carriers
\cite{Basov5}. 

\subsubsection{Transport results}

The above expressions were applied \cite{Ashk01}, using a minimal set of
realistic parameters. The transport coefficients were calculated as a
function of $T$, for five stoichiometries of p-type cuprates, ranging
from $n^p=0.8$, corresponding to the UD regime, to $n^p=0.4$,
corresponding to the OD regime. GTC auxiliary spectral functions, and
zero-energy $T$-dependent scattering rates were used, and also
impurity-scattering $T$-independent terms. The stripon band was modeled
by a ``rectangular" spectral function $A^p$ of width $\omega^p$, for
which was a value of $\sim 0.02\;$eV was found to be consistent with
experiment. 

The TEP results depend strongly on $n^p$, and reproduce very well
\cite{Ashk01} the doping-dependent experimental behavior
\cite{Fisher,Tanaka}, which has been used to determine stoichiometry.
The low-$T$ stripon-like result saturates at $T \gta 200\;$K to a
constant $(k_{_{\rm B}} / {\rm e}) \ln{[n^p/(1-n^p)]}$ \cite{Ashk01}.
But then a crossover starts towards a QE-like linear behavior, for
which a negative slope is predicted \cite{Ashk01}, in agreement with
experiment \cite{Tanaka}. The stripon term $S^p(T)$ vanishes for $n^p =
\half$, and then the TEP is determined by negative-slope QE term
$S^p(T)$, corresponding experimentally to $x = x_c \simeq 0.19$. Also
the results for the Hall coefficients \cite{Ashk01} reproduce very well
the experimental behavior \cite{Kubo,Hwang}. The approximately linear
increase of $n_{_{\rm H}}$ with $T$ is due to its crossover from a
low-$T$ stripon-like $n^p_{_{\rm H}}$ towards a QE-like $n^q_{_{\rm
H}}$. The signs of $n^p_{_{\rm H}}$ and $n^q_{_{\rm H}}$ are the same,
both determined by the nature of the one-band current ${\bf j}$ of the
averaged homogeneous planes. Thus $n_{_{\rm H}}$ is {\it not} expected to
change sign with $T$ within the assumed one-band approximation [leading
to Eq.~(10)]. 

The calculated $T$ dependence of $\rho$ \cite{Ashk01} is linear at high
$T$, flattening to (stripon-like) ``superlinearity'' at low $T$ (for all
stoichiometries). Experimentally \cite{Takagi}, the low-$T$ behavior
crosses over from ``sublinearity'' in the UD regime (and a non-metallic
upturn at lower $T$ if $T_c$ is low enough), to superlinearity in the OD
regime. As is discussed below, in relation to optical conductivity, the
sublinear behavior is the effect of reduced scattering rate in the PG
state. The low-$T$ upturn results from the localization in this state,
discussed above. The crossover to superlinear behavior (predicted here)
in the OD regime is a natural consequence of the disappearance of the PG
with increasing $x$ (see Fig.~3). The linear $T$-dependence of $\rho$
persists to low-$T$ for $x \sim 0.19$ \cite{Oh}, which is due to quantum
criticality \cite{Marel6} close to the QCP (discussed above). In the
critical region there is only one energy scale, which is the
temperature, resulting in MFL-type behavior \cite{Varma}. 

The TEP in n-type cuprates is normally expected \cite{Ashk01} to behave
similarly to the TEP in p-type cuprates, but with an opposite sign and
slope. Results for NCCO \cite{Takeda} show such behavior for low doping
levels, but in SC doping levels the high-$T$ slope of $S$ is changing
from positive to negative, and its behavior resembles that of OD p-type
cuprates. This led \cite{Ashk01} to the suggestion that NCCO may be not
a real n-type cuprate, its stripons being based on holon states (as in
p-type cuprates). 

More recent results on the n-type infinite-layer SLCO \cite{Williams} do
show TEP results for an SC cuprate which have the opposite sign and
slope than those calculated for p-type cuprates \cite{Ashk01}, as is
expected for real n-type cuprates (thus with stripons based on excession
states). It is likely that the sign of the TEP slope in NCCO changes
with doping because the one-band approximation, leading to Eq.~(10),
becomes invalid. This is indicated by ARPES \cite{Armitage2}, and by the
change with temperature of the sign of the Hall constant of NCCO
\cite{Takeda}, at the stoichiometries where the sign of the slope of the
TEP has changed (see above). Also, in YBCO, the contribution of an
additional band of the chains' carriers results in an almost zero
high-$T$ slope of the TEP \cite{Fisher}, rather than the negative slope
predicted by the GTC \cite{Ashk01}. 

\subsection{Superfluid density}

The Uemura's plots \cite{Uemura} give information about the effective
density of SC pairs $n_s^*$ through Eq.~(4). One of the mysteries of the
cuprates has been the boomerang-type behavior \cite{Niedermayer} of
these plots for $x\gta 0.19$. The author connected this behavior, as
early 1994 \cite{Ashk94}, with the fact that the (presently called)
stripon band passes through half filling. As was mentioned above, the
effect of pairing is the hybridization between stripon and QE pairs 
\cite{Ashk03}. The determination of $\lambda$ in the $\mu SR$
measurements \cite{Uemura} is through DC in a magnetic field, and as was
determined in Eqs.~(16)--(18), and the discussion following them, DC
below $T_c$ is stripon like for lightly UD, optimal (OPT), and OD
stoichiometries. 

Thus $n_s^*$ determined from $\lambda$, through $\mu SR$ measurements
\cite{Uemura}, approximately corresponds, in this regime, to the density
of stripon pairs. As was mentioned above, the stripon band is half full
for $x = x_c \simeq 0.19$, and consequently $n_s^*$ is maximal around
this stoichiometry, being determined (for p-type cuprates) by the
density of hole-like stripon pairs for $x < x_c$, and of particle-like
stripon pairs for  $x > x_c$. This result is not changed by the
intrinsic heterogeneity \cite{Ashk04} for $x\lta 0.19$, discussed above;
$T_c$ is determined there, through Eq.~(3), by the average value of
$n_s^*$, determined by measuring the penetration depth which
\cite{Tajima} is larger than the sizes of the heterogenous regions
\cite{Davis2}. 

In the crossover to the heavily UD ($x < x_0^{\prime}$) regime, the
behavior of low-$T$ DC crosses over, as was discussed above, from being
stripon-like to being QE-like. Thus, the effective superfluid density
$n_{s,{\rm eff}}$ (derived from the measured $\lambda$) becomes 
dominated, for $x < x_0^{\prime}$, by the QE contribution to the
superfluid. As was discussed above, the glassy structure in PG-like SC
regions results in the formation of localization gaps for QE states
around the antinodal points, thus preventing their contribution to the
superfluid when their localization gaps are, approximately, greater than
their pairing gaps. Since the fraction of the QE states with
localization gaps, which are small enough to contribute to the
superfluid, decreases as $x \to x_0$ (and $n_s^* \to 0$), the
contribution of QE's to the superfluid decreases then {\it faster} than
its density $n_s^*$. Thus, it is reasonable to assume in this regime an
approximate expression of the form $n_{s,{\rm eff}} \propto
(n_s^*)^{\beta}$, where $\beta > 1$. Consequently, Eqs.~(3) and (4)
yield in this regime $T_c \propto n_s^* \propto n_{s,{\rm
eff}}^{1/\beta}$. And indeed, penetration-depth measurements in the
heavily UD regime \cite{Zuev} reveal such a behavior with $\beta = 2.3
\pm 0.4$. 

\section{Optical Conductivity within the $\lowercase{\pmb{ab}\/}$ Plane}

\subsection{One-band formalism}

It is assumed that the optical conductivity, within the $ab$ plane, is
dominantly contributed (within the frequency range of interest) by one
band \cite{Macridin} of the homogeneous planes. Thus the relevant
Hamiltonian is expressed as: 
\begin{equation}
{\cal H} = \sum_{{\bf k}, \sigma} \epsilon_{\rm b}({\bf k})
d_{\sigma}^{\dagger}({\bf k}) d_{\sigma}^{\dagg}({\bf k}) + {\cal
H}_{\rm int}. 
\end{equation}
It includes a ``bare band'' [$\epsilon_{\rm b}({\bf k})$] one-particle
term, and a two-particle interaction term ${\cal H}_{\rm int}$. ${\cal
H}$ determines electron velocities ${\bf v}({\bf k})$, in terms of which
the electrical current operator (due to it) is expressed as: 
\begin{equation}
{\hat {\bf j}} \cong -{\rm e}\sum_{{\bf k}, \sigma} {\bf v}({\bf k})
d_{\sigma}^{\dagger}({\bf k}) d_{\sigma}^{\dagg}({\bf k}). 
\end{equation}

The electron creation and annihilation operators
$d_{\sigma}^{\dagger}({\bf k})$, and $d_{\sigma}^{\dagg}({\bf k})$, are
expressed in terms of the auxiliary-space QE and convoluted
stripon--svivon operators. Consequently the $\omega$-dependent
electrical current, in the presence of an electric field ${\bf
E}(\omega)$, can be expressed as a sum of a QE (${\bf j}^q$), a
stripon--svivon (${\bf j}^p$), and a mixed (${\bf j}^{qp}$)term: 
\begin{equation}
{\bf j}(\omega) = \langle {\hat {\bf j}}(\omega) \rangle = {\bf
j}^q(\omega) + {\bf j}^p(\omega) + {\bf j}^{qp}(\omega). 
\end{equation}
Since the svivons carry no charge, their effect on ${\bf j}^p$ is
through field-independent spin occupation factors. On the other hand,
${\bf j}^{qp}$ involves field-induced transitions between QE and
stripon--svivon states, and thus svivon excitations. Since
$A^{\zeta}(\omega=0) =0$, the contribution of ${\bf j}^{qp}$ to Eq.~(21)
vanishes for $\omega \to 0$, and Eq.~(10) can be used for DC. 

\subsection{The f--sum rule}

If the entire frequency spectrum were considered, the real part of the
electrons' contribution to the optical conductivity $\sigma(\omega) =
{\bf j}(\omega) / {\bf E}(\omega)$ should obey
\cite{Kubo1,Tinkham,Norman,Hanke,Marel3} the f--sum rule
$\int_0^{\infty} \Re\sigma(\omega)d\omega = \pi n {\rm e}^2/2m_e$, where
$n$ and $m_e$ are the electrons' (total) density and (unrenormalized)
mass. By considering only the one-band contribution of ${\cal H}$ in
Eq.~(19), one obtains the partial f--sum rule
\cite{Kubo1,Tinkham,Norman,Hanke,Marel3} ($V$ is the volume): 
\begin{eqnarray}
\int_0^{\infty} \Re\sigma(\omega)d\omega &=& {\pi {\rm e}^2 \over
2V}\sum_{{\bf k}, \sigma} { \langle d_{\sigma}^{\dagger}({\bf k})
d_{\sigma}^{\dagg}({\bf k}) \rangle \over m_{\rm b}({\bf k})}, \\ {1
\over m_{\rm b}({\bf k})} &\equiv& {1 \over \hbar^2}
{\partial^2\epsilon_{\rm b}({\bf k}) \over \partial {\bf k}^2}. 
\end{eqnarray}

The effective mass $m_{\rm b}({\bf k})$, appearing in Eq.~(22), is
within the bare band in ${\cal H}$ [Eq.~(19)]. The effect of the ${\cal
H}_{\rm int}$ term there is through the occupation factors $\langle
d_{\sigma}^{\dagger}({\bf k}) d_{\sigma}^{\dagg}({\bf k}) \rangle$ in
Eq.~(22). Since a large-$U$-limit method is applied here for ${\cal
H}_{\rm int}$, an optical determination of $\langle
d_{\sigma}^{\dagger}({\bf k}) d_{\sigma}^{\dagg}({\bf k}) \rangle$
requires the consideration of transitions including states in the lower,
as well as the upper Hubbard band (both in p-type cuprates and in n-type
cuprates), which means the inclusion of energies $\omega \gta 2\;$eV in
the integral in Eq.~(22). This conclusion is supported by the XAS
results \cite{Sawatzky} for the position of the chemical potential in p-
and n-type cuprates. 

The optical conductivity can be determined experimentally by reflectance
measurements and a Kramers--Kronig analysis \cite{Tanner1,Marel3}, and
the integral in Eq.~(22) can be carried out up to an experimental cutoff
frequency $\omega_{\rm co}$. This procedure has to be modified in the SC
state \cite{Tanner1,Marel3}, where the superfluid spectral weight,
contributing to the integral, comes from a $\delta$-function term in
$\Re\sigma(\omega)$ \cite{Tinkham}. This spectral weight equals
$\omega_{\rm ps}^2/8$, where $\omega_{\rm ps}$ is the superfluid plasma
frequency, and it can be determined \cite{Tanner1,Marel3} by
ellipsometry measurements, including the evaluation of the real part of
of the dielectric function $\epsilon(\omega)$, and the application of
the relation: $\Re\epsilon(\omega) = \epsilon(\infty) - (\omega_{\rm
ps}/\omega)^2$. 

This derivation of the superfluid plasma frequency provides an optical
method to determine the value of the parameter $\rho_s$ [see Eq.~(4)],
through the relation $\rho_s = \omega_{\rm ps}^2$. There is a question
concerning the relation between this optically-derived value and the
value of $\rho_s$ derived through measurements of the penetration depth
$\lambda$ in a magnetic field, by methods like $\mu SR$. The carrier
masses, determined by the two methods, are not identical; while the
optical measurements yield the bare-band mass $m_{\rm b}({\bf k})$,
appearing in Eq.~(22), carriers dynamics in a magnetic field depends on
the effective mass $m_s^*$ of the pairs. The considerable reduction of
$m_s^*$ compared to the stripon's mass is the driving force of the GTC
pairing mechanism \cite{Ashk03}. 

And indeed, it turns out \cite{Tajima} that the $ab$ plane penetration
depth determined optically, is about twice the value determined by $\mu
SR$. Also, as was discussed above, the stripons' band is half full for
$x\simeq 0.19$, and $m_s^*$ changes there from being hole like to being
particle like, resulting in a boomerang-type behavior of $\rho_s$ in the
OD regime (thus changing from increasing to decreasing with $x$), in
agreement with $\mu SR$ results \cite{Niedermayer}. On the other hand
$m_{\rm b}({\bf k})$, which determines the optical $\rho_s$, corresponds
to the bare mass of the averaged homogeneous CuO$_2$ planes, which is
hole like, and does not pass through half filling within the SC doping
range. Thus the optically determined $\rho_s$ is expected to rise with
$x$, and have no boomerang-type behavior, as has been observed
\cite{Timusk1}. 

\subsection{``Violations'' of the f--sum rule}

In ordinary SC's \cite{Tinkham}, the formation of an SC gap in
$\Re\sigma(\omega)$ is followed by the transfer of spectral weight of
magnitude $\omega_{\rm ps}^2/8$ from it to the $\delta$-function term.
Thus, the value of the integral in Eq.~(22) (including the $\omega_{\rm
ps}^2/8$ contribution in the SC state) is not expected then to change
between the normal and the SC state, when $\omega_{\rm co}$ is taken
sufficiently above the gap energy. Ellipsometric measurements in p-type
cuprates \cite{Basov3,Marel1,Bontemps,Homes2} confirm such behaviour in
the OD regime, while for UD and OPT stoichiometries, this behavior was
found to be ``violated'' even when $\omega_{\rm co}$ values above
$2\;$eV were used. This ``violation'' points to the transfer of spectral
weight below $T_c$ from energies $\gta 2\;$eV to the vicinity of
$E_{_{\rm F}}$ (its validity is confirmed in a recent debate about it
\cite{Boris}). 

There have been theoretical suggestions trying to explain this transfer
of spectral weight as being due to a mechanism of pairing from a non-FL
normal state, to an ``FL SC'' state \cite{Norman}, due to pairing phase
fluctuations \cite{Hanke}, or due to pairing via spin fluctuations
within the nearly AF FL model \cite{Carbotte1}. But the high energy
scale $\gta 2\;$eV involved is hard to understand unless it is assumed
that the spectral weight is transferred from {\it both} the lower and
the upper Hubbard bands \cite{Ashk03,Ashk04} (thus beyond the range of
applicability of the $t$--$t^{\prime}$--$J$ model). As was discussed
above, the spectral weight in the vicinity of $E_{_{\rm F}}$ is
increasing with $x$ at the expense of the weight in the upper and lower
Hubbard bands far from it, resulting in an increase in the itineracy of
the carriers. The change in $x$ is provided by doping atoms out of the
CuO$_2$ planes. These atoms contribute electronic states close to
$E_{_{\rm F}}$ [and thus also contribute to $\sigma(\omega)$], from
which charge is transferred to the CuO$_2$ planes. 

The UD and OPT stoichiometries, where the transfer of spectral weight at
$T_c$ has been observed \cite{Basov3,Marel1,Bontemps,Homes2}, are those
where the SC transition is from the PG state (see Fig.~3). As was
discussed above, the transition there is due the establishment of phase
coherence of existing localized pairs, turning them into a superfluid.
There are two possible mechanisms (or their combination) for an
accompanying transfer of spectral weight from the Hubbard bands to the
vicinity of $E_{_{\rm F}}$, both driven by the free energy gain in the
SC state. The first one is that this transfer of spectral weight is
associated with the increased itineracy of the pairs in the superfluid.
The second mechanism is that since the free energy gain due to SC is
determined (like $T_c$) by the phase stiffness, which scales [see
Eqs.~(3) and (4)] with the density of pairs within the CuO$_2$ planes,
it drives further charge transfer from the doped atoms to the planes
below $T_c$. This results in transfer of spectral weight from the
Hubbard bands to the vicinity of $E_{_{\rm F}}$, as if $x$ were
increased. On the other hand, in the OD regime the SC transition is more
BCS-like, due to pairing of electrons in an FL state, and such a
transfer of spectral weight is expected less, if at all, within both of
the above mechanisms. 

\subsection{Optical carriers around optimal stoichiometry}

\subsubsection{Contributions to the effective density}

An experimental study of the optical conductivity, and of the partial
f--sum rule, as a function $\omega_{\rm co}$, for a eight different
cases of p-type cuprates near OPT doping, was carried out by Tanner {\it
et al.} \cite{Tanner1}. A typical curve of $\Re\sigma(\omega)$, for
different temperatures, is shown in Fig.~4. The electron mass $m_e$ was
chosen for $m_{\rm b}({\bf k})$ in Eq.~(22), which is unjustified beside
being a working assumption (it is not clear at this point whether an
effective mass derived in an LDA-based calculation would be an
appropriate choice either). 

\begin{figure}[t] 
\begin{center}
\includegraphics[width=3.25in]{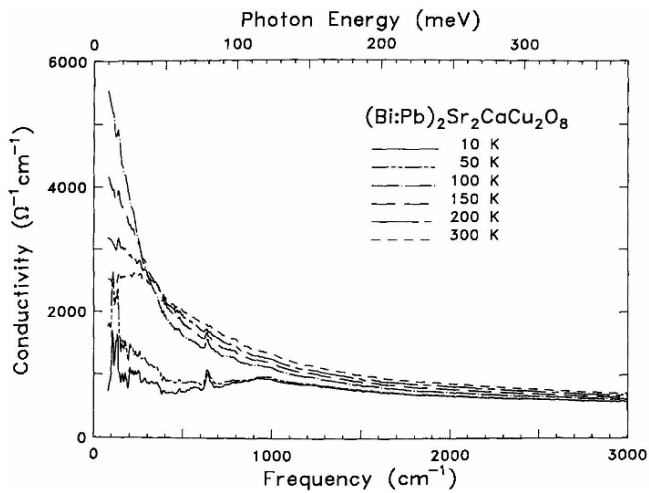}
\end{center}
\caption{Optical conductivity $\Re\sigma(\omega)$, presented by Tanner
{\it et al.} \cite{Tanner1}, at six temperatures.} 
\label{fig4}
\end{figure}

\begin{table*}[t] 
\caption{Effective number of carriers per copper in a variety of cuprate
materials, presented by Tanner {\it et al.} \cite{Tanner1}, . $N_{\rm
eff}$ is the total doping-induced carrier density, $N_{\rm s}$ the 
number of superfluid carriers, and $N_{\rm D}$ the number of Drude 
carriers. Note that the number of coppers in YBa$_2$Cu$_3$O$_7$ was 
taken as 2 with polarization (Pol.) along the $a$-axis and as 3 for 
polarization along the $b$-axis.}
\begin{center}
\includegraphics[width=7.00in]{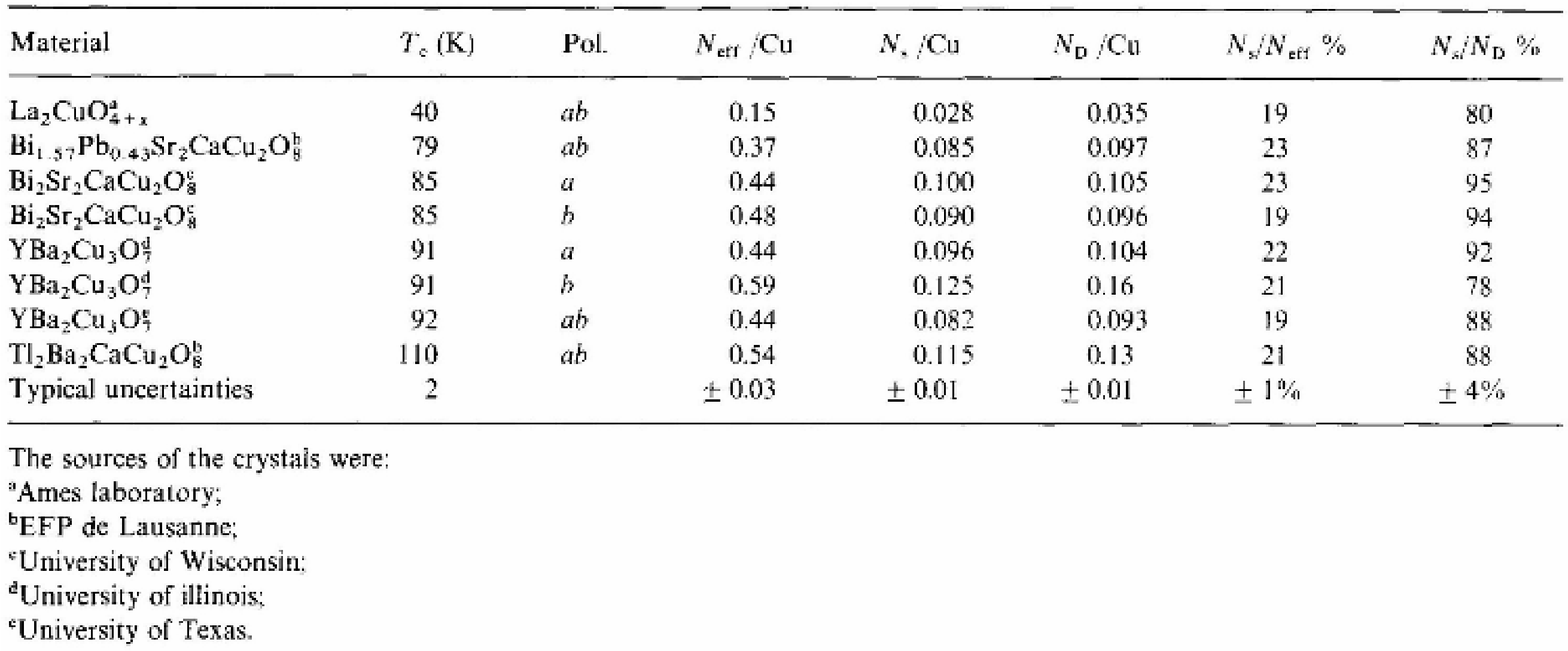}
\end{center}
\label{tab1}
\end{table*}

A value $\omega_{\rm co} \simeq 12000\;{\rm cm}^{-1}$ was chosen
\cite{Tanner1} in order to avoid the effect of the ``charge transfer''
band, and count mainly the carriers in the conduction band. As can be
concluded, {\it e.g.}, from the measured position of the chemical
potential \cite{Sawatzky}, such a choice of $\omega_{\rm co}$ in
Eq.~(22) does {\it not} count the contribution of most of the
excession-based states of the upper Hubbard band (ignored in the
$t$--$t^{\prime}$--$J$ model), while it does count the contribution of
most of the holon-based states of the (Zhang-Rice-type) lower Hubbard
band (considered in the $t$--$t^{\prime}$--$J$ model). Also the
contribution of states of the doped atoms (out of the CuO$_2$ planes) is
counted if they are close enough to $E_{_{\rm F}}$. In YBCO this
corresponds to the chains states, and they contribute considerably in
measurements with polarization in the chains direction ($b$). 

Optical carriers' densities derived through this experimental analysis
\cite{Tanner1} are presented in Table I. One derived quantity is the
effective number $N_{\rm eff}/{\rm Cu}$ of carriers per Cu atom at
$T\simeq 100\;$K (above $T_c$). Most values are found to be around
0.4--0.5, except for a lower value of 0.15 for monolayer LCO, and a
higher value of 0.59 (due to the contribution of the chains' carriers)
for YBCO with the polarization taken in the $b$ direction. It may be
misleading to connect this $N_{\rm eff}$ with an actual number of
carriers, because the contribution of the upper-Hubbard-band states is
not integrated in Eq.~(22). These states are essential to determine the
number of carriers contributed by the QE's, which form bands {\it
combining} states of the lower and the upper Hubbard bands. On the other
hand the number of carriers contributed by stripons could be well
described just within the frame of the integrated lower-Hubbard-band
states. 

Two other types of carriers' densities \cite{Tanner1}, presented in Table 
I, may have more physical significance. The first one is the
number $N_{\rm s}/{\rm Cu}$ of the superfluid carriers per Cu atom,
which was determined (as was explained above) from the $\delta$-function
term in $\sigma(\omega)$ at $T\simeq 10\;$K. The second one is the
number $N_{\rm D}/{\rm Cu}$ of Drude carriers per Cu atom (at $T\simeq
100\;$K). It was determined by fitting $\sigma(\omega)$ to the
contributions of a number of oscillators, including a Drude oscillator
at zero frequency, and Lorentzian oscillators at higher frequencies.
$N_{\rm D}$ was obtained by integrating over the Drude contribution,
which introduces (for $T > T_c$) the major low-$\omega$ contribution to
$\sigma(\omega)$ \cite{Tanner1} in Fig.~4. 

Both $N_{\rm D}$ and $N_{\rm s}$ are related to the number of carriers
involved in DC conductivity, which as was shown in Eq.~(16), and the
discussion following it, is dominantly stripon-like at low $T$, for the
stoichiometry studied. Thus $N_{\rm D}$ and $N_{\rm s}$ approximately
correspond to the number of stripon ``holes'', where $N_{\rm s}$ has
somewhat larger QE contribution due to the effect of the gap on
Eq.~(12). The contribution to the current in Eq.~(21) of ${\bf
j}^{qp}(\omega)$ (due to transitions between QE and stripon--svivon
states) could be neglected in the treatment of DC above, but it does
result in much of the non-Drude (often called mid-IR) contribution to
$\sigma(\omega)$ \cite{Tanner1} in Fig.~4. The dressing by phonons,
discussed above, is reflected in signatures of their structure there.
Also are included in this term QE contributions which (as was discussed
above) are largely ``blocked'' at low $\omega$ and $T$ by the QE
chemical potential gradient [see Eq.~(13)], satisfying there
${\bfnabla}\mu^q \cong -{\rm e}{\bf E}$. 

\subsubsection{Stripon-like carriers}

Thus, it is not surprising that the ratios $N_{\rm s}/N_{\rm D}$
\cite{Tanner1}, presented in Table I, are close to one (ranging between
0.87 and 0.95, except for values around 0.80 for LCO and YBCO with
polarization in the $b$ direction). If all the Drude carriers were
paired, the somewhat larger QE contribution to $N_{\rm s}$ would have
yielded for it a larger value than $N_{\rm D}$. However, since as was
discussed above, there are some unpaired carriers for OPT stoichiometry,
in the PG-like heterogenous regions [as can be seen in $\sigma(\omega)$
below $T_c$ \cite{Tanner1} in Fig.~4], one gets somewhat smaller $N_{\rm
s}$ than $N_{\rm D}$. The lower $N_{\rm s}/N_{\rm D}$ ratio in YBCO is
due to the contribution of the chain carriers, whose pairing could be
approximately regarded as induced by proximity \cite{Kresin}. Since the
QE contribution to the carriers' density above $T_c$ increases with $T$,
the lower $T_c$ in LCO is consistent with a lower $N_{\rm s}/N_{\rm D}$
ratio in it. 

As was discussed above, the TEP results \cite{Ashk01} indicate a
half-filled stripon band for $x = x_c \simeq 0.19$. Thus, for the
striped structure shown in Fig.~1, the OPT stoichiometry (see Fig.~3),
around which the measurements by Tanner {\it et al.} \cite{Tanner1} were
carried out, corresponds to a number of $N_{\rm con}/{\rm Cu} \simeq
0.125 \times 0.17/0.19 = 0.11$ stripon hole carriers per Cu atom. The
values of $N_{\rm s}/{\rm Cu}$ and $N_{\rm D}/{\rm Cu}$ for most of the
cases \cite{Tanner1} in Table I are quite close to 0.10, except for
smaller values for LCO, and larger values for YBCO with polarization in
the $b$ direction. 

This overall agreement is surprisingly good, considering the fact that
$m_e$ was used for $m_{\rm b}({\bf k})$ in Eq.~(22). The deviation in
YBCO is understood due to the contribution of the chains' carriers. The
deviation in LCO could be because $m_{\rm b}({\bf k})/m_e$ is
significantly larger than one there. Pavarini {\it et al.}
\cite{Andersen} have shown that when the parameters for a one-band
approximation are derived from first-principles calculations, cuprates
with larger maximal $T_c$ have larger $t^{\prime}$ hopping parameters,
and thus smaller $m_{\rm b}({\bf k})$. So having larger $m_{\rm b}({\bf
k})/m_e$ for LCO than for the other cuprates studied (in agreement with
Ref.~\cite{Andersen}) is consistent with its considerably lower $T_c$.
Also for TBCCO, which has higher $T_c$ than the other cuprates studied,
its somewhat larger values of $N_{\rm s}/{\rm Cu}$ and $N_{\rm D}/{\rm
Cu}$, than of the other cuprates presented in Table I, corresponds to
its smaller $m_{\rm b}({\bf k})/m_e$. 

\subsubsection{Tanner's law and its resolution}

The puzzling result of Tanner {\it et al.} \cite{Tanner1} (known as
Tanner's law) is that, for all the cases presented in Table I, $N_{\rm
s}/N_{\rm eff}$ ranges between 0.19 and 0.23, which means that that
$N_{\rm eff}$ equals 4-5 times the number of stripon carriers. If the
integration through Eq.~(22) were extended to include the contribution
of the upper-Hubbard-band states, {\it but} omitting the contribution of
bands not included in ${\cal H}$ in Eq.~(19), than the number of
carriers per Cu atom would have been $1-x$ (since the bare band is short
by $x/2$ from being half full for each spin state). For OPT
stoichiometry this corresponds to about 0.83 carriers per Cu atom,
which is greater than the $(N_{\rm eff}/{\rm Cu})$ values measured
\cite{Tanner1} on the basis of partial integration (omitting the
upper-Hubbard-band states), and presented in Table I. It is, however,
unrealistic to count {\it just} the contribution of the conduction-band
carriers if the integration range in Eq.~(22) is extended to include the
upper-Hubbard-band states, since they overlap with other bands. 

An analysis of $\sigma(\omega)$ on the basis of the lower-Hubbard-band
states {\it alone} (counted in Ref.~\cite{Tanner1}) could be made, in
analogy to the Kohn-Sham approach \cite{Kohn}, by replacing ${\cal H}$
in Eq.~(19) by an effective Hamiltonian of small-$U$ carriers, of the
same spin symmetry as the carriers in the real system. In this system an
effective time- and spin-dependent single-particle potential is
introduced in order to simulate (at least approximately) the many-body
effects (causing the existence of the upper-Hubbard-band) occurring in
the real system. As was shown in Eq.~(16), and the discussion following
it, such a many-body effect is that at low $\omega$ (thus DC) and $T$
the contribution of QE's to transport is largely blocked around OPT 
stoichiometry by the QE chemical potential gradient. 

In the effective system the large-$U$ constraint, which results in
different chemical potentials for QE's and stripons, doesn't exist, and
a single-particle potential has to replace it as a mechanism for
blocking the contribution of QE's, but not of stripons, to conductivity
at low $T$ and $\omega$. This effective potential has to be dynamical to
simulate the effect of the dynamical stripe-like inhomogeneities, and
its time average should maintain the translational symmetry of the
CuO$_2$ planes, as in the bare band in Eq.~(19). In the effective
system, as in the real one, the stripons correspond (see Fig.~1) to
about a quarter of the relevant orbitals of the Cu atoms in the planes,
while the QE's to the remaining three quarters. 

Such an effective dynamical potential exists, and it induces an
effective (dynamical) SDW, where minigaps are created between the states
corresponding to three quarters of the Cu atoms' states, thus blocking
their contribution to conductivity at low $T$ and $\omega$, while those
corresponding to the remaining quarter of Cu atoms' states continue
contributing to conductivity. It also induces a dynamical charge
transfer between these two types of Cu atoms. However, this effect is
minor since the bare QE and stripon states are strongly renormalized,
through the coupling between them \cite{Ashk01,Ashk03,Ashk04} [see
Eq.~(2)]. Thus the (dynamical) charge transfer between Cu atoms in the
AF and the charged stripes is considerably smaller than what would be
concluded from the occupation of stripon and QE states, if they were
(wrongly) approximated by their bare states. Such a reduced charge
transfer is estimated from the neutron-scattering \cite{Tran1} and STM
\cite{Davis1} results. 

In the effective system, interband transitions across the minigaps would
recover in higher $\omega$ and $T$ the contribution to conductivity of
the three quarters of the orbitals, blocked at low $T$ and $\omega$, and
the f--sum rule would indicate a total number of carriers $N_{\rm eff}$
which is about 4 times the number of (stripon) carriers, contributing to
transport at low $T$ and $\omega$. The contribution of carriers residing
in the inter-planar layers, discussed above, alters the factor 4 to a
somewhat higher factor, but this is partly compensated by the minor
(dynamical) charge transfer between planar Cu atoms, mentioned above,
and Tanner's law \cite{Tanner1} is obtained. 

\subsection{Optical carriers in the heavily UD regime}

\subsubsection{Low-energy excitations}

An experimental study of the optical conductivity, together with DC
transport, for LSCO and YBCO in the heavily UD regime, through
$x_0^{\prime}$, $x_0$, and the AF phase boundary, was carried out by
Padilla {\it et al.} \cite{Basov5}. Results at low $T$ (7K or just above
$T_c$) and $x < x_0^{\prime}$ (thus in the regime where transport is
dominated by QE's on the nodal FS arcs) show \cite{Basov5} that the
Drude term in $\sigma(\omega)$ is fairly separated there from the mid-IR
contribution to it. In this case, $\omega_{\rm co}$ in the integration
in Eq.~(22) could be chosen low enough to include, approximately, {\it
only} the Drude term in $\sigma(\omega)$. 

\begin{figure}[t] 
\begin{center}
\includegraphics[width=3.25in]{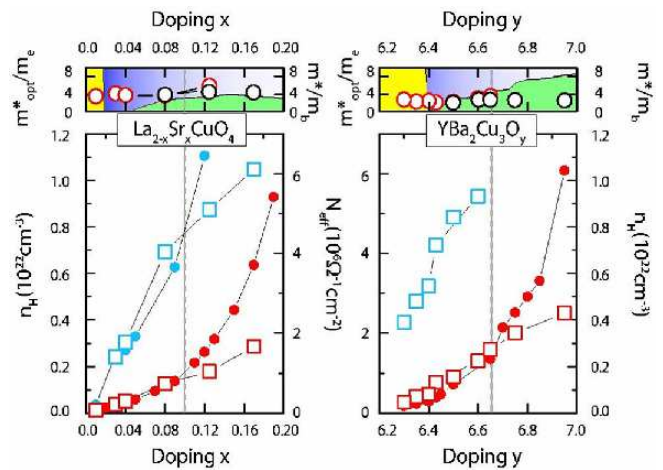}
\end{center}
\caption{(Color online) Results presented by Padilla {\it et al.}
\cite{Basov5}. Top panels show the optical effective mass determined by
the methods described in the text. Red symbols are the mass determined
using a combination of optics and transport (left axis labels), and the
black symbols represent the mass determined within the extended Drude
model \cite{Puchkov,Basov} (right labels). The low temperature long
range AF ordered phase is denoted by the yellow shaded area and SC
region by the green area. Bottom panels show both the effective spectral
weight from optics data (red and blue open squares) and the number of
holes as determined by transport (red and blue dots).} 
\label{fig5}
\end{figure}

Then, similarly to the above analysis of Tanner's law, ${\cal H}$ in
Eq.~(19) could be replaced (in analogy to the Kohn-Sham approach
\cite{Kohn}) by an effective Hamiltonian, for which the Drude term would
be {\it the only} contribution to $\sigma(\omega)$, and the effective
periodic potential would yield carriers with the effective mass
$m_{_{\rm D}}$ of the Drude carriers. Thus, by choosing such
$\omega_{\rm co}$, this effective system would yield: $\rho_{_{\rm D}}/8
= \int_0^{\omega_{\rm co}} \Re\sigma(\omega)d\omega = \pi n_{_{\rm D}}
{\rm e}^2/2m_{_{\rm D}}$, where $n_{_{\rm D}}$ is the density of the
Drude carriers. 

Results obtained by Padilla {\it et al.} \cite{Basov5} are presented in
Fig.~5. The density $n_{_{\rm D}}$ was estimated by measuring the
low-$T$ Hall constant (thus assuming $n_{_{\rm D}} = n_{_{\rm H}}$),
which by Eq.~(17), and the discussion following it, corresponds in this
regime primarily to the QE's of the nodal FS arcs, as well. The
effective mass $m_{_{\rm D}}$ of these carriers was estimated by
combining the optical and the Hall results, presented, respectively, by
red empty squares and dots in the bottom panels in Fig.~5 ($N_{\rm eff}$
there is $\propto \rho_{_{\rm D}}$). 

The evaluated values of $m_{_{\rm D}}$ are presented as red empty
circles in the top panels of Fig.~5, and they turn out \cite{Basov5} to
be $\sim 4m_e$ in LSCO, and $\sim 2m_e$ in YBCO, with almost no change
over the range $0.01 \lta x < x_0^{\prime} \simeq 0.09$. The larger mass
in LSCO is consistent with the results of Tanner {\it et al.}
\cite{Tanner1} discussed above (see Table I). The constant value found
for $m_{_{\rm D}}$ in this regime confirms the GTC scenario that no
divergence of the effective mass is occurring for $x \to 0$
\cite{Basov5}, but that carriers are doped within the Hubbard gap, and
their density is increasing with doping. These carriers become localized
at low $T$ in the low-$x$ regime, and form an FL in the high-$x$ regime.
The low-$x$ and high-$x$ limits are separated by the SC phase, or by a
QCP in the case that SC is suppressed (see Fig.~3). 

Padilla {\it et al.} \cite{Basov5} found that when $x$ is increased
above $x_0^{\prime}$, the overlap between the energy ranges of the Drude
and the mid-IR contributions to $\sigma(\omega)$ is growing, and thus
the above analysis in terms of a separate Drude term is becoming
inappropriate. Nevertheless, a trend can still be observed \cite{Basov5}
in Fig.~5 that the low-$T$ $\rho_{_{\rm D}}$ continues its increase with
$x$ at about the same rate as for $x < x_0^{\prime}$, while the rate of
increase of the low-$T$ $n_{_{\rm H}}$ grows substantially for $x >
x_0^{\prime}$. This may indicate an increase in the effective mass of
the carriers for $x > x_0^{\prime}$, confirming the GTC prediction of a
crossover in the type of carriers dominating low-$T$ transport, between
QE's on the nodal arcs, and the higher-effective-mass stripons. 

\subsubsection{High-energy excitations}

The study by Padilla {\it et al.} \cite{Basov5} included also the
determination of the effective mass $m_{\rm eff}$, on the basis of a
high-$\omega$ $\rho_{\rm eff}$, and a high-$T$ $n_{\rm eff}$. The first
was obtained by integrating on $\Re\sigma(\omega)$ in Eq.~(22) up to
$\omega_{\rm co} \simeq 12000\;{\rm cm}^{-1}$ (in order to avoid the
effect of the ``charge transfer'' band, as was discussed above
\cite{Tanner1}). The effective density ($n_{\rm eff} = n_{_{\rm H}}$)
was obtained by measuring $n_{_{\rm H}}$ in the high-$T$ Hall-constant
plateau, reached at about $800$--$900\;$K \cite{Basov5} (thus above
$T^*$). These optical and the Hall results are presented, respectively,
by blue empty squares and dots in the bottom panels in Fig.~5 ($N_{\rm
eff}$ there is $\propto \rho_{\rm eff}$). The obtained values
\cite{Basov5} of $m_{\rm eff}$ for $x < x_0^{\prime}$ in LSCO, are about
the same as the low-$T$ $m_{_{\rm D}}$ in this regime, discussed above
(presented as red empty circles in the top panels of Fig.~5). 

Both the high-$T$ plateau in $n_{_{\rm H}}$ [see Eq.~(17)], and the
high-$\omega$ $\rho_{\rm eff}$, in the heavily UD ($x < x_0^{\prime}$)
regime, correspond within the GTC mainly to the contribution of QE's.
But while at low $T$ and $\omega$, the QE's determining transport are
those on the nodal FS arcs, at high $T$ (above $T^*$) and $\omega$ {\it
all} the QE's in the conduction band become available for transport, due
to the closing of the PG. Thus, the the result: $m_{\rm eff} \simeq
m_{_{\rm D}}$ \cite{Basov5} indicates that about the same effective mass
is relevant for both cases. Temperatures in the $800$--$900\;$K range
still correspond  to a lower energy than that determined by the AF
exchange coupling in the cuprates \cite{Anisimov1,Munoz}, and thus both
$m_{_{\rm D}}$ and $m_{\rm eff}$ are expected to be determined mainly by
$t^{\prime}$ processes \cite{Ashk03}, which do not compete with the
effect of AF exchange. 
 
An analysis of the relation between $\rho_{\rm eff}$ and $\rho_{_{\rm
D}}$, in the $x < x_0^{\prime}$ regime, can be carried out similarly to
the analysis of Tanner's law above. But here the contribution of the
carriers to transport, except for the QE's on the nodal FS arcs, is
blocked, at low $T$ and $\omega$, by (real) gaps, and becomes available
when the wide frequencies range is considered. Thus, a different factor
is expected for $\rho_{\rm eff} / \rho_{_{\rm D}}$ in this regime, than
the 4-5 factor \cite{Tanner1} obtained above in Tanner's law around the
OPT regime. And indeed, Padilla {\it et al.} \cite{Basov5} got
$\rho_{\rm eff} / \rho_{_{\rm D}} \simeq 6$-7, for $x < x_0^{\prime}$,
and a crossover to Tanner's 4-5 factor, as $x$ is increased above
$x_0^{\prime}$, as is expected from the crossover to a stripon-dominated
low-$T$ transport in the rest of the doping regime. 

\begin{figure}[t] 
\begin{center}
\includegraphics[width=3.25in]{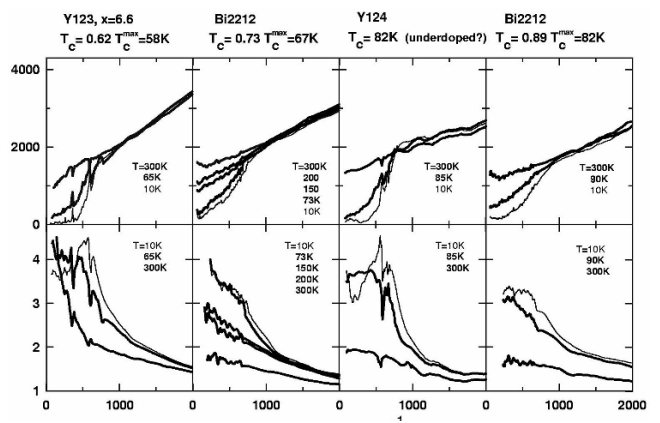}
\end{center}
\caption{The frequency dependent scattering rate, top row, and the mass
renormalization, bottom row, presented by Puchkov {\it et al.}
\cite{Puchkov}, for a series of UD cuprate SC's. The scattering rate
curves are essentially temperature independent above $1000\;{\rm
cm}^{-1}$, but develop a depression at low temperature and low
frequencies. The effective mass is enhanced at low temperature and low
frequencies.} 
\label{fig6}
\end{figure}

Padilla {\it et al.} \cite{Basov5} also applied the ``extended Drude
model'' \cite{Puchkov,Basov}:
\begin{eqnarray} 
{m^*(\omega) \over m_{\rm b}} &=& -{\omega_{\rm pn}^2 \over 4\pi\omega}
\Im\bigg[{1 \over \sigma(\omega)} \bigg],\\ \omega_{\rm pn}^2 &=&
\rho_{_{\rm D}} = {4 \pi {\rm e}^2 n_{_{\rm D}} \over m_{_{\rm D}}},
\end{eqnarray} 
($\omega_{\rm pn}$ is the normal-state plasma frequency) to estimate the
mass renormalization $m^*(\omega=0) / m_{\rm b}$, in LSCO and YBCO,
about room temperature, for $x$ below and above $x_0^{\prime}$. The
obtained values are presented as black empty circles in the top panels
of Fig.~5. The results for $m^* / m_{\rm b}$ are close to the values
(presented in red-circles) obtained for $m_{_{\rm D}} / m_e$ for $x <
x_0^{\prime}$, at low $T$ and $\omega$. 

As is demonstrated in the bottom row of Fig.~6, for results of Puchkov
{\it et al.} \cite{Puchkov} for UD cuprates, such an estimate of $m^* /
m_{\rm b}$ involves an extrapolation to $\omega=0$ from the mid-IR
range, reflecting again a dominant contribution of QE's, both for $x <
x_0^{\prime}$ and $x > x_0^{\prime}$. Thus the agreement with the
effective mass of the QE's on the nodal FS arcs, which are dominant for
$x < x_0^{\prime}$ at low $T$, is expected. This mass renormalization
reflects the omission of the contribution of $t$ processes 
\cite{Ashk03} (which, unlike $t^{\prime}$ processes \cite{Ashk03} {\it
do} compete with the effect of AF exchange). Thus the $m^*(\omega) /
m_{\rm b}$ ratio is increased for $\omega$ smaller than the energy
effect of AF exchange \cite{Anisimov1,Munoz}, as is seen in Fig.~6. This
ratio is greater in LSCO than in YBCO (see Fig.~5) because of the
smaller $t^{\prime}/t$ ratio for LSCO \cite{Andersen}. 

\subsection{In--gap states}

\subsubsection{Low-$T$ unpaired carriers and their nature}

As was discussed after Eq.~(7), in the UD regime, where the
$\bar\epsilon^p({\bf k})^2$ term in Eq.~(6) is considerably smaller than
the $\Delta^p({\bf k})^2$ term at $T \to 0$, the Bogoliubov
transformation dictates \cite{Ashk04} an approximate half filling of the
stripon band. Such half filling (namely $n^p = \half$), corresponds by
the TEP results \cite{Ashk01} to a lightly OD stoichiometry of $x = x_c
\simeq 0.19$. Thus, it was concluded \cite{Ashk04} that only a part of
the stripons are paired at $T \to 0$, in the UD regime. 

The occupation of stripon states can be regarded as consisting of $n^p$
stripon ``particles'', and $(1-n^p)$ stripon ``holes'', per stripon
state. The $T \to 0$ stripon-pairing scenario, within the UD regime, can
be approximately described as a situation where each paired stripon
``hole'' is ``coupled'' with a stripon ``particle'' (thus yielding lower
and upper Bogoliubov bands, each of approximately equal contributions of
stripon ``holes'' or ``particles''). Within this scenario, there are
unpaired states of stripon ``holes'', which are not ``coupled'' to
stripon ``particles'', and {\it vice versa}. 

In p-type cuprates, a linear approximation, expressing the dependence of
the number/state of stripon ``holes'' on $x$, yields: $1-n^p \simeq
x/2x_c$. Since the $T \to 0$ number/state of paired stripon ``holes''
drops from $1-n^p$, at $x=x_c$, to zero at $x=x_0$, a linear
approximation in $x$ would yield for it: $(1-n^p)(x-x_0)/(x_c-x_0)$.
Thus, the $T \to 0$ number/state of unpaired stripon ``holes'' would
then be approximated as: $(1-n^p)[1 - (x-x_0)/(x_c-x_0)] =
(1-n^p)(x_c-x)/(x_c-x_0)$. 

As was shown in Eq.~(16), and the discussion following it, the major
contribution to $\sigma_{ab}(\omega)$, for low $T$ and $\omega$, in the
OPT and lightly UD regime (corresponding to $x_c > x \gta
x_0^{\prime\prime} \simeq 0.13$ \cite{Basov5}), is due to stripon
``holes'', both above and below $T_c$. Thus (at low-$\omega$ in p-type
cuprates in this regime), a density of approximately $n_{\rm con}
\propto (1-n^p)$ carriers exists in $\sigma_{ab}(\omega)$ just above
$T_{\rm pair}$. At $T \to 0$, a part $n_{\rm prd}$ of this density is of
paired carriers, while the other part, $n_{\rm unp}$, is of carriers
which remain unpaired, and become localized below a temperature $T_{\rm
loc}$. Using the above linear interpolation in $x$, one can express: 
\begin{eqnarray}
n_{\rm con} &\propto& 1-n^p \simeq {x \over 2x_c}, \nonumber \\ n_{\rm
prd} &\simeq& n_{\rm con}\Big({x-x_0 \over x_c -x_0}\Big), \nonumber \\
n_{\rm unp} &\simeq& n_{\rm con}\Big({x_c-x \over x_c -x_0}\Big), \\
0.13 &\simeq& x_0^{\prime\prime} \lta x < x_c \simeq 0.19. \nonumber 
\end{eqnarray}

As was discussed above, the QE contribution to the Drude term in
$\sigma_{ab}(\omega)$ (as carriers' density $n_{\rm con}$ above $T_{\rm
pair}$, and $n_{\rm unp}$ below it), and to the $\delta$-function
superfluid term in it (as carriers' density $n_{\rm prd}$), is growing
when $x$ is decreased. The relative QE contribution to the Drude term
above $T_{\rm pair}$ is growing when $T$ is increased. The low-$T$
unpaired carriers (of density $n_{\rm unp}$) become dominated by QE's on
the nodal FS arcs, for $x < x_0^{\prime} \simeq 0.09$. The crossover
between the regimes of stripon- and QE-dominated low-$T$ transport
occurs for $x_0^{\prime} < x < x_0^{\prime\prime}$ \cite{Basov5}. 

The observed behavior \cite{Tanner1,Lupi1,Lupi2,Kim} of the
low-$\omega$ $\sigma_{ab}(\omega)$ below $T_{\rm pair}$ (both in the SC
and PG states) confirms the above predictions. This is viewed in the
effect of the $n_{\rm unp}$ unpaired carriers, appearing as a low
frequency Drude-like term \cite{Tanner1,Lupi1,Lupi2,Kim} (see Figs.~4, 
7, and 8). The width of this Drude term is smaller than that of the
normal-state Drude term \cite{Tanner1}, and it is further decreased
\cite{Tanner1,Lupi2} when the temperature is decreased, which will be
shown below to be a consequence of the GTC. This Drude term turns into a
low-$\omega$ peak in $\sigma_{ab}(\omega)$ for $T < T_{\rm loc}$
\cite{Tanner1,Lupi2}, as is expected for localized carriers. Also the
spectral weight within this Drude-like term agrees with the trend
predicted in Eq.~(26) for $n_{\rm unp}$, being small for the OPT
stoichiometry \cite{Tanner1}, where $x$ is only a little below $x_c$,
and increasing \cite{Lupi2} as $x$ is decreased within the UD regime
(see below). 

\subsubsection{The contribution of ${\bf j}^{qp}(\omega)$} 
 
Another contribution to $\sigma_{ab}(\omega)$, which is in agreement
with the GTC predictions, is the one due to transitions between QE and
stripon--svivon states \cite{Lupi1,Lupi2,Kim,Carbotte2,Timusk}, and its
evolution with $T$ for $0 < T < T_{\rm pair}$. Since the stripe-like
inhomogeneities are generated by the Bose condensation of the svivons,
and their structure is determined by the details of the of the svivon
spectrum, around its energy minimum (see Fig.~2), attempts to interpret
the structure contributed to $\sigma_{ab}(\omega)$ as a signature of
stripes \cite{Lupi1,Lupi2} are consistent with the present approach.
Since the interval of svivon energies involved is between
$+\bar\epsilon^{\zeta}({\bf k}_{\rm min})$ and
$-\bar\epsilon^{\zeta}({\bf k}_{\rm min})$, and thus its width is equal
by Eq.~(9) to the resonance-mode energy, attempts to interpret the
contributed optical structure in terms of spin excitations, and
specifically the resonance mode \cite{Carbotte2,Timusk}, are also
consistent with the present approach. 

The energies involved in transitions between QE and stripon--svivon
states include the energy difference between an occupied QE, and an
unoccupied stripon state, or {\it vice versa}, plus or minus a svivon
energy. As was discussed above, the stripon bandwidth is $\sim 0.02\;$eV,
in the normal state, and smaller in a pairing state due to the nature of
the expression for Bogoliubov quasiparticle energies, in Eq.~(6). The
QE's, on the other hand, have a wide energy spectrum around $E_{_{\rm
F}}$. The svivon spectrum is sketched in Fig.~2; its bottom is $\sim
0.02\;$eV below zero, and its top is few tenths of an eV above zero. So
these transitions contribute to $\sigma_{ab}(\omega)$ a wide, almost
featureless, spectrum, forming a major part of its mid-IR background
\cite{Tanner1} (see Fig.~4). An important role in these transitions is
played by svivons around their energy minimum at ${\bf k}_0$ where, by
Eq.~(1), the $\cosh{(\xi_{{\bf k}})}$ and $\sinh{(\xi_{{\bf k}})}$
factors appearing in the scattering Hamiltonian in Eq.~(2) are large. As
was discussed above, they are excited during transitions between
stripons and QE's around the antinodal points. 

When the stripon and QE pairing gaps open below $T_{\rm pair}$, the
width of svivon states around their minimum at ${\bf k}_0$ decreases, as
was discussed before Eq.~(9). In an analogous manner to the
spectroscopic results in Eqs. (8) and (9), one gets that transitions
between QE states on one side of the pairing gap, and stripon--svivon
states on its other side, result in a contribution to
$\sigma_{ab}(\omega)$ around the gap-edge energy. This contribution
narrows down below $T_c$ to a peak of the width $\gta W_{\rm peak}$,
given in Eq.~(9), which equals the resonance-mode energy. Note, however,
that in the heavily OD regime, the dispersion in the stripon Bogoliubov
band becomes larger than $W_{\rm peak}$, resulting in the smearing of
the peak due to the ${\bf k}$-integration taking place when the optical
spectrum is derived. Such a peak of the predicted width has been
observed by Hwang {\it et al.} \cite{Timusk}, over a wide range of
doping, and was found to ``disappear'' in the heavily OD regime, which,
as predicted above, is due to smearing. 

\begin{figure}[t] 
\begin{center}
\includegraphics[width=3.25in]{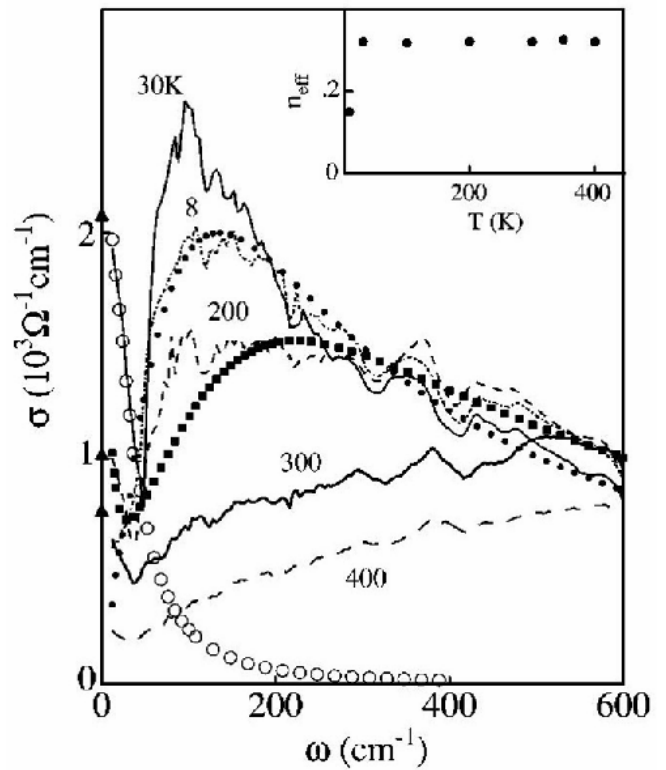}
\end{center}
\caption{Far-IR range of $\Re\sigma(\omega)$ of a BSCO film, presented
by Lupi {\it et al.} \cite{Lupi1}. The open circles are the best fit to
a normal Drude term, here proposed only for the 30~K curve. The dots and
squares are best fits to proposed model curves \cite{Lupi1} to data at 8
and 200~K, respectively. The triangles on the ordinate axis represent
the values of $\sigma_{_{\rm DC}}$ measured at the same temperatures.
The inset shows the spectral weight calculated through the partial
f--sum rule.} 
\label{fig7}
\end{figure}

\subsubsection{Evolution of the states with doping}
 
The existence of unpaired stripon states within the pairing gap, for $x
< x_c$, causes (as was discussed above) an increase in the width of this
peak above $W_{\rm peak}$ (as is noticed in Ref.~\cite{Timusk} in the UD
regime). The inclusion of transitions between QE and unpaired
stripon--svivon states results in a wider peak in $\sigma_{ab}(\omega)$,
centered within the energy range of the optical ``gap'' [which, as will
be discussed below, does not appear as a gap in $\sigma_{ab}(\omega)$].
Furthermore, as was discussed above, the unpaired stripons [see
Eq.~(26)] introduce also a Drude-like term, turning for $T < T_{\rm
loc}$ into a peak in $\sigma_{ab}(\omega)$, which merges with the one
due to transitions between QE and stripon--svivon states into one peak,
within the range of the optical ``gap''. The appearance of this merged
peak could be regarded as a signature of the glassy (checkerboard)
structure intertwined with the localization of the stripon carriers, and
the creation of an associated gap. 

\begin{figure}[t] 
\begin{center}
\includegraphics[width=3.25in]{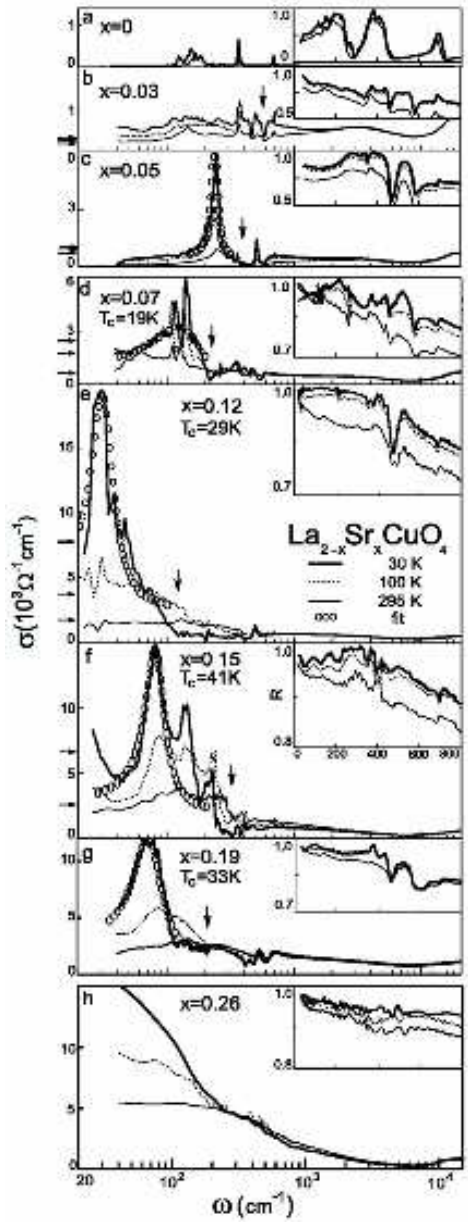}
\end{center}
\caption{Optical conductivity in the $ab$ plane,
$\Re\sigma_{ab}(\omega)$, presented by Lucarelli {\it et al.}
\cite{Lupi2}, for La$_{2-x}$Sr$_x$CuO$_4$ with increasing hole doping,
from top to bottom. Raw reflectivity data are shown in the insets for
the far-IR range. Horizontal arrows mark the DC conductivity measured at
the same temperature as the spectrum plotted with corresponding symbols.
The fits and vertical arrows are explained in Ref.~\cite{Lupi2}.} 
\label{fig8}
\end{figure}

Measurements of the evolution of this structure in
$\sigma_{ab}(\omega)$, as $T$ is lowered through $T^*$ and $T_c$, were
presented in Refs. \cite{Lupi1,Lupi2,Kim}. The results of Lupi {\it et
al.} \cite{Lupi1} on single-layer BSCO are presented in Fig.~7. They
show a Drude term at very low energies, due to the density $n_{\rm unp}$
of unpaired carriers below $T_{\rm pair}$ given in Eq.~(26). They also
show a term due to transitions between QE and stripon--svivon states,
which is moved from the mid-IR range, and narrows down, as $T$ is
lowered, into a peak overlapping energies within the range of the
optical gap, as is suggested above. 

The results of Lucarelli {\it et al.} \cite{Lupi2} on LSCO for eight
doping levels in the range $0 < x < 0.26$ are presented in Fig.~8. They
show a Drude normal behavior for $x = 0.26$, while for $x = 0.19$, 0.15,
they show the evolution of a peak due to transitions between QE and
stripon--svivon states, similarly to the one shown in Fig.~7 for BSCO
\cite{Lupi1}. A low-energy Drude term, due unpaired carriers below
$T_{\rm pair}$ (as in Fig.~7) is observed in Fig.~8 for $x = 0.15$.
Sharp peaks due to phonon modes are observed too. For $x = 0.12$, this
Drude term turns (due to localization in the glassy structure) into a
peak, merging (as was suggested above) with the other peak (due to
transitions between QE and stripon--svivon states), and the
center of the combined peak shifts to lower energy. A similar low-$T$
merged peak, within the SC optical gap, is observed also in Fig.~4
\cite{Tanner1}. 

For $x = 0.07$ and $x = 0.05$ \cite{Lupi2} this peak in Fig.~8 shifts to
a higher energy, with the decrease of $x$, which is consistent with the
increase of the localization gap. This peak turns into a wider
background for $x= 0.03$ \cite{Lupi2}, consistently with the change of
the structure into that of diagonal stripes, and it almost disappears
for $x=0$. Padilla {\it et al.} \cite{Basov6} studied the evolution of
this structure, and the phonon modes in it, in the range $0 \le x \le
0.08$, where $x$ passes through $x_0 \simeq 0.05$. Lee {\it et al.}
\cite{Basov4} found a similar behavior in the very low doping regime in
YBCO. 

Kim {\it et al.} \cite{Kim} studied LSCO for $x = 0.11$, 0.09, 0,07,
0.063, and also obtained in this regime a combined peak, as in Fig.~8
(interpreted here to be related to the glassy structure), which is
shifting to a higher energy, with the decrease of $x$. At $T = 300$K
\cite{Kim} they extrapolated a low-energy Drude term, due to the
contribution of QE's on the nodal FS arcs, discussed above
\cite{Basov5}. This Drude term is of a different nature than that due to
the stripon contribution to $n_{\rm unp}$ [given in Eq.~(26)], which is
observed at higher values of $x$. 

\section{Optical Conductivity in the $\lowercase{\pmb{c}\/}$ Direction}

Optical conductivity out of the CuO$_2$ planes, and specifically in the
$c$ direction, cannot be discussed within the one-band Hamiltonian of
Eq.~(19), and inter-planar orbitals are involved. The normal-state DC
conductivity in the $c$ direction is coherent in cases like OD YBCO 
\cite{Homes4}, due to the chains orbitals, and incoherent in other
cases. SC in the $c$ direction has been often attributed
\cite{Shibauchi,Basov1,Basov2} to Josephson tunneling between the
CuO$_2$ planes \cite{Ambegaokar,Lawrence}. This scenario is {\it not}
distinct from the description of the $c$-direction SC as bulk SC within
the dirty limit. Both scenarios give the same temperature dependence of
the $c$-direction penetration depth \cite{Shibauchi,Basov1}. 

The inter-planar layers may be too thin to be described as ``metallic''
or ``insulating''. Since the orbitals of the SC electrons of the CuO$_2$
planes are hybridized with inter-planar orbitals, the Cooper pairs of
the planes do penetrate the inter-planar layers, and in cases that these
layers contribute many carriers around $E_{_{\rm F}}$, proximity-induced
pairing is expected to exist there \cite{Kresin}. Upon crossing to the
OD regime \cite{Homes4}, coherence is $c$ direction can be established
in the normal state, and the dirty-limit scenario looks more natural
than the Josephson scenario, but still the $c$-direction SC features are
not significantly altered. 

The SC transition is followed by the establishment of a $c$-direction
SC plasmon mode \cite{Marel3,Dulic2,Dordevic1}, which is referred to,
within the Josephson tunneling scenario, as a Josephson plasma
resonance. But it is also expected within the GTC pairing mechanism,
discussed above \cite{Ashk03}, due to the hybridization of orbitals of
the inter-planar layers with the CuO$_2$-plane QE states, resulting in
some inter-plane pair hopping. Effects of the SC transition on the
$c$-direction optical conductivity include a similar ``violation'' of
the f--sum rule \cite{Basov3,Marel4}, as in the $ab$-plane optical
conductivity, due to transfer of spectral weight from energies $\gta
2\;$eV, which was explained above within the GTC. 

Another $c$-direction optical effect is the transfer of spectral weight
from the SC gap to the mid- or far-IR range, which has been identified
\cite{Marel3,Marel2,Dulic1,Marel4,Marel5,Dordevic2} as the signature of
a $c$-axis collective mode. In bilayer structures it was attributed to a
transverse out-of-phase bilayer plasmon, related to ``excitons'' first
considered by Legget \cite{Legget}, due to relative phase fluctuations
of the condensates formed in two different bands. The observation of a
similar effect in monolayer LSCO \cite{Marel4} suggests that the LaSrO
layers, where the doped atoms reside, may introduce enough states close
to $E_{_{\rm F}}$ to provide the second band needed for such transverse
plasmons to be formed. This collective mode was found
\cite{Marel4,Dordevic2} to be well defined already in the PG state. This
is consistent with the GTC under which the pairs already exist in this
state, and even though they lack long-range phase coherence, short-range
effects over the distance between close planes/layers, required for this
mode to exist, are expected to be present. 

\section{Optical Scattering Rates}

Using the extended Drude model \cite{Puchkov,Basov}, the optical
scattering rate can be expressed as [see Eq.~(25)]: 
\begin{equation}
{1 \over \tau(\omega)} = {\omega_{\rm pn}^2 \over 4\pi} \Re\bigg[{1
\over \sigma(\omega)} \bigg] 
\end{equation}
(thus the carriers density dependence of the conductivity is eliminated
to obtain the scattering rate). The $ab$-plane and $c$-direction
scattering rates, $\tau_{ab}(\omega)^{-1}$ and $\tau_c(\omega)^{-1}$,
behave differently in the cuprates. Quantum criticality \cite{Marel6}
close to the QCP in the normal state (see Fig.~3) results in the
existence of a critical region where only one energy scale, which is the
temperature, exists. Consequently an MFL-type behavior \cite{Varma} is
obtained there (thus linearity in $\omega$) for $\tau_{ab}(\omega)^{-1}$
and related quantities \cite{Timusk2}, and further optical quantities
behave critically \cite{Marel6}. 

As was shown in Eq.~(16), and the discussion following it, the
conductivity in the $ab$ plane, $\sigma_{ab}(\omega)$, is dominantly
stripon-like for low $T$ and $\omega$ for $x > x_0^{\prime\prime}$ [see
Eq.~(26)]. Consequently, $\tau_{ab}(\omega)^{-1}$ is determined in this
regime by the scattering of stripons, through ${\cal H}^{\prime}$ [see
Eq.~(2)] into QE and svivon states \cite{Ashk01,Ashk03}. As was
discussed before Eq.~(9), the main contribution to such scattering comes
\cite{Ashk04} from  QE states around the antinodal points, and svivon
states around their energy minimum at ${\bf k}_0$ (see Fig.~2), where,
by Eq.~(1), the $\cosh{(\xi_{{\bf k}})}$ and $\sinh{(\xi_{{\bf k}})}$
factors are large. Below $T_{\rm pair}$ a QE pairing gap opens around
the antinodal points, and as was discussed above, all the QE states
there either become paired, or have a localization gap, for $T \to 0$.
This results in a reduction, below $T_{\rm pair}$, of
$\tau_{ab}(\omega)^{-1}$, for $\omega$ within the pairing gap. This
reduction becomes drastic at low $T$, when the width of the gap-edge
peak becomes small (because of the exclusion of such scattering), and
the QE gap approaches its $T \to 0$ value. 

Since the width of the Drude term in $\sigma_{ab}(\omega)$ is
$\tau_{ab}^{-1}$, its decrease below $T_{\rm pair}$ results in a smaller
Drude width for the unpaired carriers there [see Eq.~(26) and the above
discussion], than for the carriers above $T_{\rm pair}$ (see Fig.~4).
Furthermore, this Drude width is expected to decrease with decreasing
temperature, in agreement with experiment \cite{Tanner1,Lupi2}. As was
discussed above, the relative contribution of QE's to the Drude term is
growing when $T$ is increasing above $T_{\rm pair}$. This results in an
increase in the width $\tau_{ab}^{-1}$ of the Drude term with $T$ above
$T_{\rm pair}$, in agreement with experiment \cite{Tanner1} (see
Fig.~4). 

For $x < x_0^{\prime\prime}$, the nature of the unpaired carriers below
$T_{\rm pair}$ crosses over from being dominantly stripon like, to being
dominated by QE's on the nodal FS arcs. However, this does {\it not}
result in an increase in their scattering rate $\tau_{ab}^{-1}$ (and
thus Drude width), because (unlike the antinodal QE's, which contribute 
significantly above  $T_{\rm pair}$) these QE's are
not scattered to stripon--svivon states, around the svivons' energy
minimum at ${\bf k}_0$ (see Fig.~2), and the contribution of other
svivons to their scattering is much smaller (as was discussed above).
And indeed, the width of the Drude term in the heavily UD regime was
found \cite{Basov5} to be pretty small and even fairly separated from
the mid-IR term, as was discussed above. 

A consequence of the existence of unpaired carriers within the gap is
that the drop in $\tau_{ab}^{-1}$ below $T_{\rm
pair}$ is not followed by an equal relative drop in the effective
density of carriers. Since these carriers contribute to DC conduction
for $T > T_{\rm loc}$, the expected effect is an {\it increase} in the
$ab$-plane DC conductivity above $T_{\rm loc}$ in the PG state. This
explains the observed \cite{Takagi} sublinear $T$-dependence of the DC
resistivity for $T_{\rm loc} < T < T^*$ in the UD regime. Also is
explained the increase in this effect on the resistivity \cite{Takagi}
when $x$ is decreased, and thus $T^*$ and the PG size are increased. 

The $c$-direction conductivity $\sigma_c(\omega)$ is, on the other hand,
determined by the QE states (with hybridization taking place between
planar states and inter-planar orbitals). Thus $\tau_c(\omega)^{-1}$ is
determined by scattering of QE's to stripon--svivon states, persisting
below $T_c$ (and thus resulting in the wide hump, as was discussed
above), and by processes within the inter-planar layers, which are
unrelated to the pairing process, and it is not expected to vary
significantly below $T_{\rm pair}$. Since and the number of unpaired
QE's drops below $T_{\rm pair}$ (while $\tau_c^{-1}$ does not vary 
significantly), a drop in $\sigma_c(\omega)$ within the gap is occurring
below $T_{\rm pair}$. The contribution to it from pairs is expected to
be transferred to the $\delta$-function term in the SC state, and to
higher frequencies in the PG state, where the pairs are localized. 

\begin{figure}[t] 
\begin{center}
\includegraphics[width=3.25in]{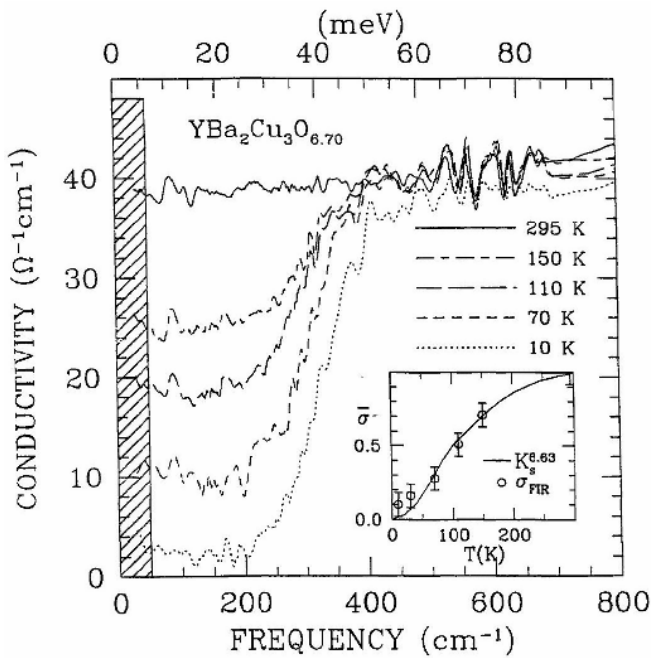}
\end{center}
\caption{The $c$-axis optical conductivity $\Re\sigma_c(\omega)$ of a UD
Y123 crystal, presented by Puchkov {\it et al.} \cite{Puchkov}. The
$c$-axis conductivity is temperature and frequency independent for
$T>T^*$, but develops a marked gap-like depression below $T^*$. As the
temperature is lowered the PG deepens. Inset: the NMR Knight shift
(normalized at 300~K) is plotted as a function of temperature for a UD
Y123 crystal. The circles show the low-frequency $c$-axis conductivity
for samples of the same doping level. The curves suggest that the Knight
shift, a conventional measure of the DOS at $E_{_{\rm F}}$, and the
$c$-axis conductivity are depressed by the same process in the PG
state.} 
\label{fig9}
\end{figure}

The contribution of the QE localization gaps to the PG (discussed
above), especially in the heavily UD regime, also contributes to the
drop in $\sigma_c(\omega)$ within the gap. The spectral weight
corresponding to these localized states is not transferred to the
$\delta$-function term, when the localization gaps (approximately)
exceed the pairing gaps, and this applies to most of the QE's when $x
\to x_0$. 

These predictions are confirmed by experiment. A sharp drop has been
observed \cite{Bonn,Puchkov} in $\tau_{ab}(\omega)^{-1}$ of the
quasiparticles below $T_c$, as can be seen in the top row of Fig.~6.
Consequently a clean-limit treatment applies in the SC state within the
$ab$ plane, while $\tau_{ab}(\omega)^{-1}$ above $T_c$ corresponds to
the intermediate scattering regime (see below) \cite{Homes5}. Also, as
can be seen in Fig.~6, the reduction in $\tau_{ab}(\omega)^{-1}$, for
$\omega$ within the gap, starts in the PG state (thus below $T^* =
T_{\rm pair}$) \cite{Puchkov,Basov}. 

On the other hand, as can be seen in Fig.~9, a gap-like depression has
been observed below $T^*$ \cite{Homes3,Puchkov} in $\sigma_c(\omega)$
within the PG energy range, with the spectral weight from it transferred
to higher energies (above $T_c$). The similarity, shown in the inset in
Fig.~9, between the gap-like behavior in $\sigma_c(\omega \to 0)$, and
the gap observed in Knight-shift results, is consistent with the GTC
prediction that this gap is in the spectrum of spin-carrying QE-like
carriers. 

\section{Homes' Law}

Homes {\it et al.} \cite{Homes1} have demonstrated that, in many SC's,
the optical quantity $\rho_s = \omega_{\rm ps}^2$ [based on a similar
expression as in Eq.~(4)], obeys approximately (thus when presented on
a log--log scale) the relation 
\begin{equation}
\rho_s \simeq 35 \sigma_{_{\rm DC}}(T_c)T_c 
\end{equation}
(both sides in the equation possess the same units). This relation,
known as Homes' law, is obeyed in the cuprates in almost the entire
doping regime, both in the $ab$-plane, and in the $c$-direction, and in
other SC's which behave according to dirty-limit approximations. In a
subsequent paper, Homes {\it et al.} \cite{Homes5} demonstrated that
this law should be valid in the dirty-limit, and in the
intermediate-scattering regime, when it applies to transport in the {\it
normal state}. The reason is that the $\rho_s/8$ spectral weight, which
condenses into the superfluid $\delta$-funcion term in $\sigma(\omega)$,
approximately scales then as the product of $\sigma_{_{\rm DC}}(T_c)$
and the SC gap, which approximately scales as $T_c$. Thus, in the regime
of validity of Homes' law, it could be expressed as: 
\begin{equation}
k_{_{\rm B}}T_c \lta \hbar \tau^{-1}(T_c),
\end{equation}

As was discussed above, $\sigma_{ab}(\omega)$ in the cuprates
corresponds to the intermediate-scattering regime in the normal state,
with most of the Drude carriers turning into a superfluid. But the drop
in $\tau_{ab}(\omega)$ below $T_c$ (seen in Fig.~6), predicted by the
GTC, turns them into clean-limit SC's. The $c$-direction conductivity
was discussed above as being described by the dirty-limit both above and
below $T_c$, and also through the Josephson tunneling scenario, which
was shown \cite{Homes1,Homes4} to result in Eq.~(28) too. Thus, the GTC
is consistent with the applicability of Homes' law in the cuprates, both
in the $ab$-plane, and the $c$-direction, and it could be expressed
there as: 
\begin{equation}
k_{_{\rm B}}T_c({\rm cuprates}) \simeq \hbar \tau^{-1}(T_c).
\end{equation} 

This expression is the consequence of the MFL behavior \cite{Varma}, due
to quantum criticality, and thus \cite{Marel6} the existence of one
energy scale, which is the temperature. The coexistence of Homes' law
with Uemura's law [see Eq.~(4)] in the UD regime implies, by Eq.~(28),
that $\sigma_{_{\rm DC}}(T_c)$ does not vary much with $x$ there (in
agreement with experiment \cite{Takagi}, except when localization starts
above $T_c$), which was shown above to be the consequence of the
decrease in $\tau_{ab}^{-1}$ [and by Eq.~(30), also in $T_c$] in the PG
state. 

Eq.~(30) connects high $T_c$ in the cuprates with a high scattering
rate, which lead also to an MIT. The existence of SC close to the
boundary of MIT's has been pointed out by Osofsky {\it et al.}
\cite{Osofsky}. Zaanen \cite{Zaanen} suggested that Homes' law implies
that $T_c$ in the cuprates is determined, approximately by the condition
that the scattering rate in the normal state is becoming at $T_c$ as
high as is permitted by the laws of quantum physics. Here it was shown
that the high scattering rate is strongly decreased below $T_c$, since
it is determined by the same interactions which determine pairing.

\section{Conclusions}

The success of the GTC in providing the understanding of a variety of
puzzling optical properties of the cuprates, in addition to its earlier
success in explaining many other anomalous properties of these
materials, strengthens the point of view that the occurrence of
high-$T_c$ SC in them requires the proximity of a Mott transition. 

It suggests the occurrence of spin-charge separation in the cuprates
\cite{Anderson}, {\it but} only due to the existence of dynamical
inhomogeneities, which provide quasi-one-dimensional structures. It also
predicts an intrinsic origin to the static nanoscale heterogeneity
observed in the UD regime \cite{Davis2}. 

Furthermore, the GTC supports the opinion that the same interactions
which play a major role in the determination of the electronic structure
of the cuprates, also primarily determine pairing, transport, and other
anomalous properties in them. 

Even though a complete rigorous first-principles proof on the validity
of the GTC is still beyond reach, due to the complexity of the global
scheme, and many of its results are still obtained on a qualitative
level, the mounting evidence on the global scope of its applicability
points very strongly to its validity for the cuprates. 
 

\end{document}